\documentclass[twocolumn,superscriptaddress,amsmath,amssymb,aps,floatfix]{revtex4-1}

\usepackage{graphicx}
\usepackage{slashed}
\usepackage{epstopdf}
\usepackage{bm}
\usepackage{hyperref}
\usepackage{graphicx,color}
\usepackage{multirow}
\usepackage{amsmath,amstext,array}
\usepackage[normalem]{ulem}

\newcolumntype{C}{>{$}c<{$}}

\begin{document}
\title{Perturbative Unitarity and the 4-Point Vertices in the Constructive Standard Model}
\author{Neil Christensen}
\email{nchris3@ilstu.edu}
\affiliation{Department of Physics, Illinois State University, Normal, IL 61790}
\date{\today}

\begin{abstract}
We find a complete set of 4-point vertices in the Constructive Standard Model (CSM).  This set is smaller than in Feynman diagrams as the CSM does not need or allow any additional 4-point vertices (or ``contact'' terms) beyond what is present in Feynman diagrams and, furthermore, it does \textit{not} need or allow a 4-point vertex for $Z Z \bar{W} W$, $W W \bar{W} \bar{W}$, $\gamma Z W \bar{W}$ or $\gamma \gamma W \bar{W}$, in addition to the already known absence of the 4-gluon vertex.  We show that with this set of 4-point vertices, perturbative unitarity is satisfied in the CSM.  Additionally, we show that many constructive diagrams are not Feynman diagrams rewritten in spinor form.  In fact, we show that there is a significant rearrangement of contributions from the diagrams in constructive calculations relative to Feynman diagrams, for some processes.  In addition to the already known or expected rearrangement in diagrams involving external photons, we also find that diagrams involving 4 vector bosons are also significantly different than their Feynman counterparts.  
\end{abstract}

\maketitle

\tableofcontents

\section{Introduction}

\begin{table}
    \centering
    \textbf{CSM 4-Point Vertices}\\
    \vspace{0.05in}
    \renewcommand{\arraystretch}{2.7}
    \begin{tabular}{|@{\hspace{5pt}}c@{\hspace{5pt}}|@{\hspace{5pt}}c@{\hspace{5pt}}|@{\hspace{5pt}}c@{\hspace{5pt}}|}
        \hline
        Particles & Coupling & Vertex \\
        \hline
         $h h h h$ & $-i\frac{3e^2m_h^2}{4M_W^2s_W^2}$ & 1 \\
         \hline
         $h h Z Z$ & $i\frac{e^2}{2M_W^2s_W^2}$ & $\lbrack\mathbf{34}\rbrack\langle\mathbf{34}\rangle$ \\
         \hline
         $h h W \bar{W}$ & $i\frac{e^2}{2M_W^2s_W^2}$ & $\lbrack\mathbf{34}\rbrack\langle\mathbf{34}\rangle$ \\
        \hline
    \end{tabular}
    \caption{Complete list of 4-point vertices in the Constructive Standard Model (CSM).  }
    \label{tab:4-point vertices}
\end{table}

Alternate techniques for calculating scattering amplitudes ``constructively", without the use of field theory or Feynman diagrams, have been gathering steam for several decades now
\cite{Parke:1986gb,Gastmans:1990xh,Dixon:1996wi,Britto:2005fq,Elvang:2013cua}.  Initially, most of the progress was for massless theories, but with \cite{Arkani-Hamed:2017jhn}, the method was extended to massive theories as well.  Following this step, full calculations in the constructive Standard Model (CSM) were begun, beginning with a complete set of 3-point vertices for the CSM \cite{Christensen:2018zcq}, and other calculations \cite{Christensen:2019mch,Ochirov:2018uyq,Franken:2019wqr,Aoude:2019tzn,Durieux:2019eor,Bachu:2019ehv,Balkin:2021dko,Baratella:2020lzz,Durieux:2020gip,Liu:2023jbq,Alves:2021rjc,Bachu:2023fjn,Liu:2022alx}. Immediately following the publication of the 3-point vertices, we set out to find the 4-point vertices of the CSM.  Our intention was to use perturbative unitarity \cite{Lee:1977yc,Appelquist:1987cf,Chivukula:2007gse,Chivukula:2007koj} to determine them, but our attempts were blocked by challenges with diagrams with internal photons \cite{Christensen:2022nja}.  This last paper found a work around using a massive photon and taking the massless limit and \cite{Lai:2023upa} showed that the $x$ factors had further structure that allowed them to be used with massless photons to obtain the correct amplitude in the case of $f \bar{f}\to f \bar{f}$, where $f$ is a fermion.  Moreover, \cite{Ema:2024vww} described the momentum shifts that enabled this procedure to work.  

With the photon diagrams solved, we set out again to find the complete set of 4-point vertices of the CSM.  This culminated in the present paper where we find all the 4-point vertices and establish perturbative unitarity in the CSM, in addition to two companion papers.  In one, we use the 4-point vertices along with the 3-point vertices to find the 4-point amplitudes in a comprehensive set of CSM processes \cite{Christensen:2024C}.  In the other, we create a C++ package called SPINAS, designed to calculate any constructive amplitude \cite{Christensen:2024B}.  Further, we use SPINAS to validate all the 4-point amplitudes in the CSM, including those presented here.

In this paper, in greater detail, we analyze the high-energy-growth terms of all the 4-point amplitudes involving longitudinal vector bosons.  We find the processes where a 4-point vertex is required in order to cancel the high-energy growth and achieve perturbative unitarity and we make a complete list of these 4-point vertices in Table~\ref{tab:4-point vertices}.  We also include the 4-Higgs vertex, for completeness and claim that this is a complete set of 4-point vertices for the Constructive Standard Model (CSM).  No other 4-point vertices (or contact terms) are required or allowed.  This includes the exclusion of 4-point vertices for $Z Z \bar{W} W$, $W W \bar{W} \bar{W}$, $\gamma Z W\bar{W}$ and $\gamma\gamma W\bar{W}$ that are present in Feynman diagrams and the already well-known absence of a four-gluon vertex.  

We further show that constructive diagrams are not always simply Feynman diagrams written in spinor notation.  In fact, in some processes, the contributions to the amplitude are significantly rearranged, so that individual constructive diagrams do not equal individual Feynman diagrams.  We find two classes where this is the case.  The first is when there are external photons or gluons.  This was already known for a small number of processes and expected in others, and we have enlarged the number of calculated processes and confirm a rearrangement in all of them.  Additionally, we show in this paper that any process with a Feynman 4-vector-boson vertex, such as $Z Z W \bar{W}$ and $W \bar{W} W \bar{W}$, also have a significant rearrangement of contributions.  This is clear to see since there is no 4-point vertex in constructive calculations.  

In order to find the 4-point vertices of the CSM, we look in the same place we find them in Feynman diagrams.  With the exception of the 4-Higgs vertex and the 4-gluon vertex, we find Feynman 4-point vertices in processes where they are required to achieve perturbative unitarity.  This occurs when a process has diagrams that grow in the limit of high energy (at tree level), which occurs with Feynman diagrams when a longitudinal vector boson is present.  This is also where we look for them in the case of the CSM.  

We can see this at a naive level by considering the contribution from the parts of a diagram.  The propagator denominator grows as the energy squared ($\mathcal{E}^2$), as usual.  The growth of the numerator depends on the number of spinors and the number of momenta in it.  Each spinor contributes $\sqrt{\mathcal{E}}$ and each momentum $\mathcal{E}$.  Therefore, we can count the expected number of each to get an idea.  A four-fermion diagram would have four spinors creating $\mathcal{E}^2$ growth that cancels with the $\mathcal{E}^2$ in the denominator if no additional momenta are present.  Therefore, we do not expect to need any 4-fermion vertices and, indeed, we find that we do not in the companion to this paper \cite{Christensen:2024C}.  The same is true if we replace two of these with Higgs bosons, which lowers the naive high-energy growth.  

However, when we begin adding vector bosons, since they need two spinors each, we increase the chance that high-energy growth will appear and require a cancellation.  For example, if we have two fermions and two vector bosons, we need six spinors, contributing $\mathcal{E}^3$ growth (divided by $\mathcal{E}^2$ from the propagator denominator), at a minimum, and requiring cancellation among the diagrams.  If we have four vector bosons, we need eight spinors, contributing $\mathcal{E}^4$ over the $\mathcal{E}^2$ from the denominator.  Again, even at the minimal level, we require high-energy cancellation.  Moreover, these diagrams with vector bosons often have extra factors of momenta increasing their high-energy growth, as we will see in detail in this paper.  

On the other hand, external photons tend to decrease the high-energy growth (relative to vector bosons) due to extra propagator denominators.  In fact, we will find in this paper that many processes involving external photons that have high-energy growth and cancellation in Feynman diagrams do not have any high-energy growth at all here in the CSM.

In this paper, we will show that the CSM fully satisfies perturbative unitarity and we will establish a complete list of 4-point vertices in the CSM.  As mentioned earlier, we include the 4-Higgs vertex for completeness, even though it is not required for perturbative unitarity of 4-point amplitudes.  It is trivially related to the Feynman 4-Higgs vertex and has been validated in \cite{Christensen:2024C}.  We break this paper up into four main sections.  In Sec.~\ref{sec:4-point:yes:yes}, we consider processes that require a 4-point vertex, namely $h h \to Z Z$ and $h h \to W \bar{W}$.  In Sec.~\ref{sec:4-point: HE growth, but no 4-point}, we consider processes that \textit{do} have high-energy growth terms that need to cancel, but do \textit{not} require or allow a 4-point vertex in the CSM, but do require one in Feynman diagrams.  This includes the processes $Z Z \to W \bar{W}$, $W \bar{W} \to W \bar{W}$.  In Sec.~\ref{sec:4-point: no:yes}, we consider processes that do \textit{not} have any high-energy growth and do \textit{not} require or allow a 4-point vertex in the CSM, but do require one in Feynman diagrams.  This includes the processes $\gamma^+ Z \to W \bar{W}$, $\gamma^+ \gamma^+ \to W \bar{W}$ and $\gamma^+ \gamma^- \to W \bar{W}$, with these choices of helicity representative of the other choices.  In Sec.~\ref{sec:4-point: no:no}, we look at a selection of the processes that do not require or allow a 4-point vertex in either the CSM or in Feynman diagrams, but which include the potential of a cancellation of high-energy-growth terms.  This includes the processes $Z Z \to Z Z$, $t \bar{t} \to W \bar{W}$, $t \bar{t} \to Z Z$ and $\bar{t} b \to Z \bar{W}$.  Although we do not include every possible process with cancellations here, we have verified that they are all perturbatively unitary.  In order to focus on some of the most significant cancellations, we use the third-generation quarks as examples since the high-energy growth terms are typically proportional to the fermion mass and, therefore, processes with the top-quark mass will pose the largest potential challenge for perturbative unitarity.  However, we find that perturbative unitarity is satisfied for the other fermions in a similar way.  In Sec.~\ref{sec:conclusions}, we conclude.  

Additionally, we have included the following appendices.  In App.~\ref{app:amplitudes}, we have included the full T- and U-channel contributions to the processes $Z Z \to W \bar{W}$ and $W \bar{W} \to W \bar{W}$, which are very long and would hurt readability of the main text.  In App.~\ref{app:ZZWW:W T HEE}, we give the complete list of high-energy-growth terms for the T-channel diagram for $Z Z \to W \bar{W}$ as an example of how many terms there are before cancellation of most of them with the U-channel diagram.  In the main body of this paper, we always combine the T- and U-channel diagrams before showing them for conciseness and readability.  In App.~\ref{app:HEG Feynman}, we show the high-energy expansion of the Feynman 4-point vertices for comparison with our CSM results.  We find that, for most of them, there is a significant difference between the CSM 4-point vertex and the Feynman 4-point vertex.  This is a nice way to see that there is a rearrangement of contributions to the amplitude.  

All calculations in this paper are performed in the center of momentum (CM) frame with 
\begin{equation}
    \mathcal{E} = \frac{\sqrt{s}}{2}
\end{equation}
where $s=(p_1+p_2)^2$ is the Mandelstahm variable and
\begin{align}
\lvert\mathbf{p}_1\rvert 
    &= \frac{1}{2}\sqrt{\frac{\left(s-(m_1+m_2)^2\right)\left(s-(m_1-m_2)^2\right)}{s}}
\\
\lvert\mathbf{p}_3\rvert 
    &= \frac{1}{2}\sqrt{\frac{\left(s-(m_3+m_4)^2\right)\left(s-(m_3-m_4)^2\right)}{s}} .
\end{align}
The third particle travels at an angle $\theta$ with respect to the first particle's direction.  Details of the spinors can be found in various places \cite{Arkani-Hamed:2017jhn,Christensen:2018zcq,Christensen:2024B}.  After calculating the spinor products for a particular spin combination, we Taylor expanded in high $\mathcal{E}$.

\section{\label{sec:4-point:yes:yes}Processes With a Constructive 4-Point Vertex}

In this section, we will consider the processes which include high-energy growth in individual diagrams and which require a 4-point vertex to cancel it in both constructive calculations and in Feynman diagrams.  There are only two such processes, namely $h h \to Z Z$ and $h h \to W \bar{W}$.

\subsection{$\mathbf{h h \to Z Z}$ and $\mathbf{h h \to W \bar{W}}$}
Both these processes have very similar details.  In fact, only the masses are changed between them.  We will describe them together, but have done them both.  The contribution to both amplitudes coming from an S-channel Higgs is
\begin{equation}
    \mathcal{M}_h = -\frac{3e^2m_h^2}{2M_W^2s_W^2}\frac{\lbrack\mathbf{34}\rbrack\langle\mathbf{34}\rangle}{(s-m_h^2)} .
\end{equation}
By inspection, we can see that there is no high-energy growth from this term.  Both the numerator and denominator grow at the same $\mathcal{E}^2$ rate, canceling at high energy.  

The contributions from a $Z$ boson in the T channel and U channel of $h h \to Z Z$ are
\begin{align}
\mathcal{M}_{TZ} &= 
    -\frac{e^{2}}{2M_W^{2} s_W^{2}\left(t-M_Z^{2}\right)} 
    \Big(
    2M_Z^{2} \langle\mathbf{3}\mathbf{4}\rangle \lbrack\mathbf{3}\mathbf{4}\rbrack 
    \nonumber\\
    &\hspace{-0.2in}+\lbrack\mathbf{3}\lvert p_{1} \rvert \mathbf{4}\rangle \lbrack\mathbf{4}\lvert p_{1} \rvert \mathbf{3}\rangle -\left(\lbrack\mathbf{3}\mathbf{4}\rbrack \lbrack\mathbf{3}\lvert p_{1} \rvert \mathbf{4}\rangle +\langle\mathbf{3}\mathbf{4}\rangle \lbrack\mathbf{4}\lvert p_{1} \rvert \mathbf{3}\rangle \right)M_Z
    \Big)
    \label{eq:hhZZ:TZ diagram}
\end{align}
and
\begin{align}
\mathcal{M}_{UZ} &= 
    -\frac{e^{2}}{2M_W^{2} s_W^{2}\left(u-M_Z^{2}\right)} 
    \Big(
    2M_Z^{2} \langle\mathbf{3}\mathbf{4}\rangle \lbrack\mathbf{3}\mathbf{4}\rbrack 
    \nonumber\\
    &\hspace{-0.2in}+\lbrack\mathbf{3}\lvert p_{1} \rvert \mathbf{4}\rangle \lbrack\mathbf{4}\lvert p_{1} \rvert \mathbf{3}\rangle +\left(\langle\mathbf{3}\mathbf{4}\rangle \lbrack\mathbf{3}\lvert p_{1} \rvert \mathbf{4}\rangle +\lbrack\mathbf{3}\mathbf{4}\rbrack \lbrack\mathbf{4}\lvert p_{1} \rvert \mathbf{3}\rangle \right)M_Z
    \Big) .
    \label{eq:hhZZ:UZ diagram}
\end{align}
For the contributions of a $W$ boson in the T and U channels of the process $h h \to W \bar{W}$, simply replace all the $M_Z$ with $M_W$.  Otherwise, the contributions look the same.

If we Taylor expand in high energy and combine the T and U channels, we are left with (in both cases)
\begin{equation}
    \mathcal{M}_{Z/W}^{(0,0)} = \frac{e^2}{M_W^2s_W^2}\mathcal{E}^2 .
\end{equation}

The 4-point vertex could potentially have contributions from $\langle\mathbf{34}\rangle^2$, $\langle\mathbf{34}\rangle\lbrack\mathbf{34}\rbrack$ and $\lbrack\mathbf{34}\rbrack^2$.  By a simple process of elimination, we find that, in order to cancel the high-energy-growth terms and achieve perturbative unitarity, the 4-point vertex is given in Table~\ref{tab:4-point vertices} and the contribution to this amplitude is given by
\begin{equation}
    \mathcal{M}_4 = \frac{e^2}{2M_W^2s_W^2}\lbrack\mathbf{34}\rbrack\langle\mathbf{34}\rangle .
\end{equation}
Taylor expanding at high energy, we are left with
\begin{equation}
    \mathcal{M}_4^{(0,0)} = -\frac{e^2}{M_W^2s_W^2}\mathcal{E}^2 .
\end{equation}
We can see that, with this, the high-energy growth cancels.  In fact, we find that the amplitude agrees at all energies.  

In App.~\ref{app:HEG Feynman:hhZZ}, we give the high-energy growth of the Feynman 4-point vertex for this process.  We can see that in this case, the high-energy growth is the same for the Feynman 4-point vertex.  Moreover, we have checked in SPINAS and it agrees with the Feynman 4-point vertex for all energies in this amplitude when squared.  We have also checked the Higgs diagram and find that it agrees with the Feynman Higgs diagram at all energies when squared.  Consequently, we find that each of these diagrams appears to be equivalent to Feynman diagrams, simply written in spinor notation.  We have checked these statements at the squared-amplitude level in SPINAS.

\section{\label{sec:4-point: HE growth, but no 4-point}Processes with No Constructive 4-Point Vertices I}
In this section, we will consider processes with four massive vector bosons.  Using Feynman diagrams, these include a 4-point vertex.  However, as we will show here, they do not involve a 4-point vertex in constructive calculations.  Although there is high-energy growth, it cancels among all the diagrams involving only 3-point vertices.  The fact that there is a 4-point Feynman vertex, but no 4-point constructive vertex, makes it immediately clear that constructive diagrams are not equivalent to Feynman diagrams and they are not simply Feynman diagrams rewritten in spinor notation.  There is a significant rearrangement of contributions, especially when a 4-point vertex is involved in one and not the other.

\subsection{$\mathbf{Z Z \to W \bar{W}}$}

The contribution to the amplitude coming from an S-channel Higgs is
\begin{equation}
    \mathcal{M}_h =
    -\frac{e^{2}}{M_W^{2} s_W^{2}}
    \frac{\langle\mathbf{1}\mathbf{2}\rangle \langle\mathbf{3}\mathbf{4}\rangle \lbrack\mathbf{1}\mathbf{2}\rbrack \lbrack\mathbf{3}\mathbf{4}\rbrack }{ \left(s-M_h^{2}\right)} .
    \label{eq:ZZWW:Higgs Amp}
\end{equation}

If we Taylor expand this in high energy, there is only one channel that contains energy growth.  It is when all the external particles have helicity 0.   We find
\begin{equation}
    \mathcal{M}_h^{(0,0,0,0)} =
    -\frac{e^{2}}{M_W^{2} s_W^{2}}\mathcal{E}^2 .
    \label{eq:ZZWW:M_h^0000}
\end{equation}
All other channels have, at most, a constant.  This agrees with the result from Feynman diagrams.

The contributions to the amplitude coming from a T- and U-channel $W$ boson are quite complicated.  We have included them in App.~\ref{app:ZZWW 4: Derivation}.  Each of these diagrams individually contributes energy growth in a large number of channels, with a maximum energy growth of $\mathcal{E}^3$.  There are no quartic high-energy growth terms in the constructive amplitude, unlike in Feynman diagrams.  All of these energy growth terms cancel except for one term.  In order to keep this section legible, we give the energy growth after combining the T and U channels.  The interested reader can see the terms for just the T channel in App.~\ref{app:ZZWW:W T HEE}.  For the U channel, they are all equal in magnitude but opposite in sign, except for the single term that we show below.  Therefore, after expansion and combining the T and U channels, we are left with
\begin{align}
    \mathcal{M}_{W}^{(0,0,0,0)} &= \frac{e^2}{M_W^2s_W^2}\mathcal{E}^2 
    \label{eq:ZZWW:0000}
\end{align}
Comparing Eqs.~(\ref{eq:ZZWW:M_h^0000}) and (\ref{eq:ZZWW:0000}), we see that all high-energy growth cancels.  There is no need for a 4-point vertex, and one is not allowed.  If a 4-point vertex were present, it would ruin perturbative unitarity at high energy.  Moreover, we have found that the amplitude with only these contributions agrees with Feynman diagrams at all energies and for a variety of masses in SPINAS.

This is rather remarkable.  It appears that the high-energy cancellation is better behaved in constructive amplitudes.  There is no quartic energy growth at all, and all the high-energy growth that is present is canceled between the T- and U-channel $W$ diagrams, with only a single quadratic energy-growth term left that cancels with the Higgs.  As we will see, the situation is similar for the four $W$-boson scattering and even better for amplitudes involving an external boson of helicity-$\pm1$.

It is remarkable for another reason.  We see that constructive diagrams are \textit{not} simply Feynman diagrams reduced to the spinor components.  Constructive amplitudes rearrange the contributions.  Even when the constructive diagrams have a resemblance to Feynman diagrams, they are not directly related for spin-1 and helicity-$\pm1$ bosons.  In this amplitude, only the Higgs contribution is the same as the Feynman-diagram contribution.  This is not only true for high energy, but for all energies.  We have checked this with SPINAS \cite{Christensen:2024B} at the squared amplitude level.

We further note that the cancellations that are present are under better control in constructive calculations.  Feynman diagrams have a quartic high-energy growth that cancels between three diagrams (the T-channel $W$, U-channel $W$ and 4-point diagrams), which is absent here.  There is no quartic energy growth in any diagram.  Furthermore, the cancellations that do take place are mostly between just two diagrams (the T-channel and U-channel diagrams), rather than the three of Feynman diagrams.

\subsection{$\mathbf{W W \to W W}$}

The Higgs boson contributes in both the T and the U channels.  The contributions to the amplitude are
\begin{align}
\mathcal{M}_{Th} &= 
    -\frac{e^2}{M_W^2s_W^2}
    \frac{\langle\mathbf{1}\mathbf{3}\rangle \langle\mathbf{2}\mathbf{4}\rangle \lbrack\mathbf{1}\mathbf{3}\rbrack \lbrack\mathbf{2}\mathbf{4}\rbrack }{ \left(t-M_h^{2}\right)} 
    \\
\mathcal{M}_{Uh} &= 
    -\frac{e^2}{M_W^2s_W^2}
    \frac{\langle\mathbf{1}\mathbf{4}\rangle \langle\mathbf{2}\mathbf{3}\rangle \lbrack\mathbf{1}\mathbf{4}\rbrack \lbrack\mathbf{2}\mathbf{3}\rbrack }{ \left(u-M_h^{2}\right)} 
    .
\end{align}

After Taylor expanding in high energy, the combined result is
\begin{equation}
\mathcal{M}_h^{(0,0,0,0)} =
    \frac{e^{2}}{M_W^{2} s_W^{2}}\mathcal{E}^2
    .
    \label{eq:WWWW:M_h^0000}
\end{equation}
The individual channels do have high-energy growth in other polarization channels, but they all cancel except this one.  This contribution to the amplitude agrees with Feynman diagrams.  

The photon contributes in the T and U channels.  Their full amplitude expressions can be found in Eqs.~(\ref{eq:WWWW:TA full}) and (\ref{eq:WWWW:UA full}).  As we can see in those expressions, the highest possible energy growth is cubic.  Although that does appear in some polarization channels for individual diagrams, after combining the T and U channels, we are left with
\begin{align}
    \mathcal{M}_{\gamma}^{(0,0,0,0)} &= -\frac{4e^2}{M_W^2}\mathcal{E}^2.
\end{align}
The energy growth in all other channels cancels.

The $Z$ boson also contributes in the T and U channels.  Their full amplitude expressions can be found in Eqs.(\ref{eq:WWWW:TZ full}) and (\ref{eq:WWWW:UZ full}).  After Taylor expanding in high energy and combining the T and U channels and only keeping the high-energy growth terms, we have
\begin{align}
    \mathcal{M}_{Z}^{(0,0,0,0)} &= 
    \frac{(3-4c_W^2)e^2}{M_W^2s_W^2}\mathcal{E}^2 .
\end{align}

Combining the contributions, we have
\begin{align}
    \mathcal{M}^{(0,0,0,0)} &=
    \frac{e^2\mathcal{E}^2}{M_W^2s_W^2}
    \left(
    1-4s_W^2+3-4c_W^2
    \right) = 0.
\end{align}
We find that all high-energy growth cancels.  We have checked every polarization channel.  

As in the case of $Z Z \to W \bar{W}$, there is no need for a 4-point vertex in this process at all and, indeed, none is allowed.  This is in contrast to Feynman rules where a 4-point vertex is required.  Once again, we see that constructive diagrams are significantly rearranged in some cases and are not Feynman diagrams written in a different form.  We have included the high-energy growth for the Feynman 4-point vertex contribution in App.~\ref{app:HEG Feynman:WWWW} to highlight the contrast.  As in the previous case, only the Higgs diagrams contribute the same as Feynman diagrams.  We have checked this with SPINAS \cite{Christensen:2024B} at the squared amplitude level for all energies.

As in the previous subsection, we again comment on the reduced dependence on precise cancellations between diagrams.  There is no quartic high-energy growth here in any diagram, and no need for its cancellation.  Moreover, the cancellations that do take place are mostly between just two diagrams at a time.  The majority of them are between the T- and U-channel diagrams with one mediator.  That is, most of the energy growth cancellations are between the photon T- and U-channel diagrams and separately between the $Z$-boson T- and U-channel diagrams, whereas in Feynman diagrams, most of these cancellations are between five diagrams (including the 4-point vertex diagram).  This should lead to better numerical stability in phase-space integrations.

\section{\label{sec:4-point: no:yes}Processes with No Constructive 4-Point Vertices II}

In this section, we consider the processes which do \textit{not} have any high-energy growth and which do \textit{not} have any 4-point vertices in the CSM, but \textit{do} have both high-energy growth and 4-point vertices in the Feynman diagram version of the SM.  We will also see that for these processes, there is a significant rearrangement of contributions and that constructive diagram are not equivalent to Feynman diagrams for these processes.  

Furthermore, the property that these amplitudes do not have any high-energy growth and, therefore, do not need any cancellations in order to satisfy perturbative unitarity extends beyond the amplitudes shown here.  We have found that it is true for any 4-point amplitude with any number of external photons or gluons \cite{Christensen:2024C}.

\subsection{$\mathbf{\gamma^+ Z \to W \bar{W}}$}
There is only one constructive diagram for this process.  It could come from either the T or the U channels.  They both give identical results, which is
\begin{align}
\mathcal{M} &=
    \frac{2e^{2}}{c_W s_W\left(t-M_W^{2}\right)\left(u-M_W^{2}\right)} 
    \Big(
    c_W^{2} \lbrack1\mathbf{4}\rbrack ^{2} \langle\mathbf{2}\mathbf{3}\rangle ^{2} 
    \nonumber\\
    &+\left(2c_W^{2}-1 \right)\lbrack1\mathbf{3}\rbrack \langle\mathbf{2}\mathbf{4}\rangle\lbrack1\mathbf{4}\rbrack \langle\mathbf{2}\mathbf{3}\rangle 
    +c_W\lbrack1\mathbf{2}\rbrack \langle\mathbf{3}\mathbf{4}\rangle\lbrack1\mathbf{4}\rbrack \langle\mathbf{2}\mathbf{3}\rangle 
    \nonumber\\
    &\hspace{-0.15in}+c_W^2\lbrack1\mathbf{3}\rbrack ^{2} \langle\mathbf{2}\mathbf{4}\rangle ^{2} 
    -c_W\lbrack1\mathbf{2}\rbrack \lbrack1\mathbf{3}\rbrack \langle\mathbf{2}\mathbf{4}\rangle \langle\mathbf{3}\mathbf{4}\rangle 
    +c_W^2\lbrack1\mathbf{2}\rbrack ^{2} \langle\mathbf{3}\mathbf{4}\rangle ^{2}
    \Big) .
    \label{eq:AZWW:W full}
\end{align}
The high-energy Taylor expansion gives no energy growth terms.  In fact, this can be seen by inspection since there are no momentum terms in the numerator.  There are four spinor products per term, each of which grows as $\mathcal{E}^4$.  But, the denominator grows at the same rate.  Thus, there is no high-energy growth.  No 4-point vertex is required to cancel any high-energy growth.  Moreover, the amplitude in Eq.~(\ref{eq:AZWW:W full}) is equal to the combined Feynman diagrams at all energies, not just at high energies, which we have checked with SPINAS.  In other words, no 4-point vertex is required \textit{or even allowed} for any energies.  This is another sign that the constructive diagrams are \textit{not} simply the Feynman diagrams expanded in terms of spinors, in general.  There is a rearrangement of contributions to the amplitude regarding the spin-1 boson contributions in some processes.  In this case, it completely eliminates the need for a 4-point vertex.  Among other things, this means that one constructive diagram in this case accounts for 3 Feynman diagrams.  

Furthermore, note that Feynman diagrams result in high-energy growth that must be cancelled between terms.  This results in large cancellations that must be carefully maintained and are a potential source of loss of precision.  In the constructive case, on the other hand, there are not high-energy growth terms with large cancellations at all.  Even within the amplitude expression, each term does not grow at high energy.  Thus, we find that constructive calculations result in better behaved numerical calculations and are under better control, especially in amplitudes, such as this one, where no high-energy growth exists at all.

Although there is no 4-point vertex here to compare with, we nevertheless give the 4-point vertex from Feynman diagrams in App.~\ref{app:HEG Feynman:AZWW} as well as its high-energy-growth terms for contrast.  We can see that there are cancellations in every polarization channel that has at least one of the vector bosons in the longitudinal mode.  In fact, in the channel where all three massive vector bosons are longitudinal, the energy growth is cubic ($\mathcal{E}^3$).  All of this is nonexistent in this constructive amplitude.

\subsection{$\mathbf{\gamma^+ \gamma^+ \to W \bar{W}}$}
The amplitude for this process is obtained from either the T- or the U-channel diagram.  They give identical results.  It is
\begin{equation}
\mathcal{M}_W = 
    \frac{2e^{2} \lbrack12\rbrack ^{2} \langle\mathbf{3}\mathbf{4}\rangle ^{2} }{\left(t-M_W^{2}\right)\left(u-M_W^{2}\right)} .
\end{equation}
It can be seen by inspection that the numerator and denominator grow at the same rate at high energy, therefore, the amplitude does not have any high-energy growth terms.  Once again, there is no need for a 4-point vertex and indeed, \textit{none is allowed}.  This amplitude agrees with Feynman diagrams for all energies, as we checked with SPINAS.  We again see that this diagram is \textit{not} a Feynman diagram written in spinor form.  There is a significant rearrangement of contributions in constructive diagrams, resulting in a much simpler result.

Again, for contrast, we show the Feynman 4-point vertex contribution in App.~\ref{app:HEG Feynman:AAWW}, along with its high-energy-growth terms.  All five longitudinal polarization channels grow at either linear or quadratic order at high energy.  These require cancellations against the other Feynman diagrams.  In the constructive case, no high-energy growth exists at all, and there is no cancellation.  As before, the calculation is much better behaved and under much better numerical control.  In fact, this process is special in that there are no cancellations at all, as there is only one term.

\subsection{$\mathbf{\gamma^+ \gamma^- \to W \bar{W}}$}
Once again, the amplitude for this process is obtained from either the T or the U channel diagram.  They give identical results.  It is
\begin{equation}
\mathcal{M}_W = 
    \frac{2e^2\left(\langle2\mathbf{4}\rangle \lbrack1\mathbf{3}\rbrack +\langle2\mathbf{3}\rangle \lbrack1\mathbf{4}\rbrack \right)^{2} }{\left(t-M_W^{2}\right)\left(u-M_W^{2}\right)} .
\end{equation}
It can be seen by inspection that the numerator and denominator grow at the same rate at high energy, therefore, the amplitude does not have any high-energy growth terms.  Once again, there is no need for a 4-point vertex and  \textit{none is allowed}.  This simple amplitude agrees with the combined Feynman diagrams for all energies, as we show with SPINAS.  Again, we see that this one diagram is equivalent to all three Feynman diagrams and there is no direct diagram-for-diagram relationship.  The contributions are rearranged.

As before, we include the Feynman 4-point vertex contribution in App.~\ref{app:HEG Feynman:AAWW}, along with its high-energy-growth terms.  As in the previous subsection, all five longitudinal polarization channels contribute high-energy growth that must be canceled against the other Feynman diagrams.  Once again, the constructive amplitude is not only simpler, but under better numerical control, since there are no high-energy-growth terms to cancel.

\section{\label{sec:4-point: no:no}Processes With No Constructive 4-Point Vertices III}

In this section, we will consider a selection of processes which do not have 4-point vertices, either in constructive calculations or in Feynman diagrams.    For the processes in this section, there is agreement between constructive and Feynman contributions for each internal particle.

\subsection{$\mathbf{Z Z \to Z Z}$}
The contribution of the Higgs boson in the S, T and U channels are
\begin{align}
\mathcal{M}_S &= 
    -\frac{e^{2} }{M_W^{2} s_W^{2}}
    \frac{\langle\mathbf{1}\mathbf{2}\rangle \langle\mathbf{3}\mathbf{4}\rangle \lbrack\mathbf{1}\mathbf{2}\rbrack \lbrack\mathbf{3}\mathbf{4}\rbrack }{ \left(s-m_h^{2}\right)} 
    \\
\mathcal{M}_T &= 
    -\frac{e^{2} }{M_W^{2} s_W^{2}}
    \frac{\langle\mathbf{1}\mathbf{3}\rangle \langle\mathbf{2}\mathbf{4}\rangle \lbrack\mathbf{1}\mathbf{3}\rbrack \lbrack\mathbf{2}\mathbf{4}\rbrack }{ \left(t-m_h^{2}\right)}    
    \\
\mathcal{M}_U &=
    -\frac{e^{2}}{M_W^{2} s_W^{2}}
    \frac{\langle\mathbf{1}\mathbf{4}\rangle \langle\mathbf{2}\mathbf{3}\rangle \lbrack\mathbf{1}\mathbf{4}\rbrack \lbrack\mathbf{2}\mathbf{3}\rbrack }{ \left(u-m_h^{2}\right)} .
\end{align}

Taylor expanding, we have for the S channel
\begin{align}
\mathcal{M}_S^{(0,0,0,0)} &= 
	-\frac{e^{2}}{ M_W^{2} s_W^{2}} \mathcal{E}^{2} .
\end{align}
The combined contribution from the T and U channels is 
\begin{align}
\mathcal{M}_{TU}^{(0,0,0,0)} &= 
	\frac{e^{2}}{ M_W^{2} s_W^{2}} \mathcal{E}^{2} .
\end{align}
which is exactly the opposite and cancels it, so there is no high-energy growth.  

This occurs, as we can see, without the need for a 4-point vertex, and none is allowed.  In this case, there is also no Feynman 4-point vertex contributing, so these cases are similar in this respect.  We have found agreement with the Feynman-diagram amplitude for all energies with SPINAS.

\subsection{$\mathbf{t \bar{t} \to W \bar{W}}$}
We will also consider processes with two fermions that have a longitudinal vector boson.  For concreteness and conciseness, we will discuss it in the context of the third-generation quarks, but the  processes with other generations of quarks or leptons are similar.

We begin with the process $t \bar{t} \to W \bar{W}$.  It has a contribution from the Higgs in the S channel, which is
\begin{equation}
    \mathcal{M}_h = 
    \frac{-e^{2} m_t}{2M_W^{2} s_W^{2}}
    \frac{ \left(\langle\mathbf{1}\mathbf{2}\rangle +\lbrack\mathbf{1}\mathbf{2}\rbrack \right)\langle\mathbf{3}\mathbf{4}\rangle\lbrack\mathbf{3}\mathbf{4}\rbrack }{ \left(s-m_h^{2}\right)} .
\end{equation}
Taylor expanding in high energy results in
\begin{align}
\mathcal{M}_{h}^{(-\frac{1}{2},-\frac{1}{2},0,0)} &= 
    \frac{e^{2} m_t}{2 M_W^{2} s_W^{2}} \mathcal{E}\\
\mathcal{M}_{h}^{(\frac{1}{2},\frac{1}{2},0,0)} &= 
    -\frac{e^{2} m_t}{2 M_W^{2} s_W^{2}} \mathcal{E} .
\end{align}

The contribution from the $b$ quark in the T channel is
\begin{equation}
    \mathcal{M}_b = \frac{-e^{2} }{M_W^{2} s_W^{2}} 
    \frac{\langle\mathbf{1}\mathbf{3}\rangle \lbrack\mathbf{2}\mathbf{4}\rbrack \left(M_W\langle\mathbf{3}\mathbf{4}\rangle -\lbrack\mathbf{3}\lvert p_{1} \rvert \mathbf{4}\rangle \right)}{ \left(t-m_b^{2}\right)} .
    \label{eq:ttWW:b T}
\end{equation}
Taylor expanding at high energy results in
\begin{align}
\mathcal{M}_{b}^{(-\frac{1}{2},-\frac{1}{2},0,0)} &= 
    -\frac{e^{2} m_t\left(1+\cos(\theta)\right)}{2 M_W^{2} s_W^{2}} \mathcal{E}\\
\mathcal{M}_{b}^{(-\frac{1}{2},\frac{1}{2},-1,0)} &= 
    -\frac{\sqrt{2}e^{2} \sin^{2}\left(\frac{\theta}{2}\right) }{M_Ws_W^{2} } \mathcal{E}\\
\mathcal{M}_{b}^{(-\frac{1}{2},\frac{1}{2},0,-1)} &= 
    \frac{e^{2} \left(1+\cos(\theta)\right)}{\sqrt{2}M_Ws_W^{2} } \mathcal{E}\\
\mathcal{M}_{b}^{(-\frac{1}{2},\frac{1}{2},0,0)} &= 
    -\frac{e^{2}\sin(\theta)}{ M_W^{2} s_W^{2}} \mathcal{E}^{2} \\
\mathcal{M}_{b}^{(-\frac{1}{2},\frac{1}{2},0,1)} &= 
    \frac{e^{2} \left(1-\cos(\theta)\right)}{\sqrt{2}M_Ws_W^{2} } \mathcal{E}\\
\mathcal{M}_{b}^{(-\frac{1}{2},\frac{1}{2},1,0)} &= 
    -\frac{e^{2} \left(1+\cos(\theta)\right)}{\sqrt{2}M_Ws_W^{2} } \mathcal{E}\\
\mathcal{M}_{b}^{(\frac{1}{2},\frac{1}{2},0,0)} &= 
    \frac{e^{2} m_t\left(1+\cos(\theta)\right)}{2 M_W^{2} s_W^{2} }\mathcal{E} .
\end{align}

The contribution from the photon in the S channel is 
\begin{align}
    \mathcal{M}_\gamma &= 
    \frac{-2e^2Q}{M_W^2s}
    \Bigg(
    M_W\left(\langle\mathbf{3}\mathbf{4}\rangle +\lbrack\mathbf{3}\mathbf{4}\rbrack \right)
    \nonumber\\
    &\times\left(\langle\mathbf{2}\mathbf{4}\rangle \lbrack\mathbf{1}\mathbf{3}\rbrack +\langle\mathbf{2}\mathbf{3}\rangle \lbrack\mathbf{1}\mathbf{4}\rbrack +\langle\mathbf{1}\mathbf{4}\rangle \lbrack\mathbf{2}\mathbf{3}\rbrack +\langle\mathbf{1}\mathbf{3}\rangle \lbrack\mathbf{2}\mathbf{4}\rbrack \right) 
    \nonumber\\
    &+\left(\langle\mathbf{3}\mathbf{4}\rangle \lbrack\mathbf{3}\mathbf{4}\rbrack \lbrack\mathbf{1}\lvert p_{3} \rvert \mathbf{2}\rangle +\langle\mathbf{3}\mathbf{4}\rangle \lbrack\mathbf{3}\mathbf{4}\rbrack \lbrack\mathbf{2}\lvert p_{3} \rvert \mathbf{1}\rangle \right)
    \Bigg) ,
    \label{eq:ttWW:A S}
\end{align}
where $Q=2/3$.
After Taylor expanding, we have
\begin{align}
\mathcal{M}_{\gamma}^{(-\frac{1}{2},-\frac{1}{2},0,0)} &= 
    \frac{2e^{2}Q m_t\cos(\theta)}{M_W^{2}} \mathcal{E}\\
\mathcal{M}_{\gamma}^{(-\frac{1}{2},\frac{1}{2},-1,0)} &= 
    \frac{\sqrt{2}e^{2}Q \left(1-\cos(\theta)\right)}{M_W} \mathcal{E}\\
\mathcal{M}_{\gamma}^{(-\frac{1}{2},\frac{1}{2},0,-1)} &= 
    -\frac{\sqrt{2}e^{2}Q \left(1+\cos(\theta)\right)}{M_W} \mathcal{E}\\
\mathcal{M}_{\gamma}^{(-\frac{1}{2},\frac{1}{2},0,0)} &= 
    \frac{2e^{2}Q\sin(\theta)}{M_W^{2}} \mathcal{E}^{2} \\
\mathcal{M}_{\gamma}^{(-\frac{1}{2},\frac{1}{2},0,1)} &= 
    \frac{\sqrt{2}e^{2} Q \left(-1+\cos(\theta)\right)}{M_W}\mathcal{E} \\
\mathcal{M}_{\gamma}^{(-\frac{1}{2},\frac{1}{2},1,0)} &= 
    \frac{\sqrt{2}e^{2}Q \left(1+\cos(\theta)\right)}{M_W} \mathcal{E}\\
\mathcal{M}_{\gamma}^{(\frac{1}{2},-\frac{1}{2},-1,0)} &= 
    -\frac{\sqrt{2}e^{2}Q \left(1+\cos(\theta)\right)}{M_W} \mathcal{E}\\
\mathcal{M}_{\gamma}^{(\frac{1}{2},-\frac{1}{2},0,-1)} &= 
    \frac{\sqrt{2}e^{2} Q \left(1-\cos(\theta)\right)}{M_W} \mathcal{E}\\
\mathcal{M}_{\gamma}^{(\frac{1}{2},-\frac{1}{2},0,0)} &= 
    \frac{2e^{2}Q\sin(\theta)}{ M_W^{2}} \mathcal{E}^{2} \\
\mathcal{M}_{\gamma}^{(\frac{1}{2},-\frac{1}{2},0,1)} &= 
    \frac{\sqrt{2}e^{2}Q \left(1+\cos(\theta)\right)}{M_W} \mathcal{E}\\
\mathcal{M}_{\gamma}^{(\frac{1}{2},-\frac{1}{2},1,0)} &= 
    \frac{\sqrt{2}e^{2}Q \left(-1+\cos(\theta)\right)}{M_W}\mathcal{E} \\
\mathcal{M}_{\gamma}^{(\frac{1}{2},\frac{1}{2},0,0)} &= 
    -\frac{2e^{2}Q m_t\cos(\theta)}{M_W^{2}} \mathcal{E} .
\end{align}

The contribution from the $Z$ boson in the S channel is
\begin{align}
\mathcal{M}_Z &= 
    \frac{-e^2}{2 M_W^2 s_W^2\left(s-M_Z^{2}\right)}
    \Bigg(
    \nonumber\\
    &2\langle\mathbf{3}\mathbf{4}\rangle \lbrack\mathbf{3}\mathbf{4}\rbrack 
    \left(g_R\lbrack\mathbf{1}\lvert p_{3} \rvert \mathbf{2}\rangle + g_L\lbrack\mathbf{2}\lvert p_{3} \rvert \mathbf{1}\rangle \right)
    \nonumber\\
    &+ m_t \left(\langle\mathbf{1}\mathbf{2}\rangle  - \lbrack\mathbf{1}\mathbf{2}\rbrack
    \right)\langle\mathbf{3}\mathbf{4}\rangle \lbrack\mathbf{3}\mathbf{4}\rbrack
    \nonumber\\
    &+2M_W    \left(\langle\mathbf{3}\mathbf{4}\rangle+\lbrack\mathbf{3}\mathbf{4}\rbrack\right)
    \Big[g_R
    \left(\langle\mathbf{2}\mathbf{4}\rangle \lbrack\mathbf{1}\mathbf{3}\rbrack +\langle\mathbf{2}\mathbf{3}\rangle \lbrack\mathbf{1}\mathbf{4}\rbrack 
    \right)
    \nonumber\\
    &+g_L \left(\langle\mathbf{1}\mathbf{4}\rangle  \lbrack\mathbf{2}\mathbf{3}\rbrack +\langle\mathbf{1}\mathbf{3}\rangle  \lbrack\mathbf{2}\mathbf{4}\rbrack  \right)
    \Big]
    \Bigg) ,
    \label{eq:ttWW:Z S}
\end{align}
where $g_L=-2Q s_W^2 + 1$ and $g_R = -2Q s_W^2$.
After Taylor expanding, we have
\begin{align}
\mathcal{M}_{Z}^{(-\frac{1}{2},-\frac{1}{2},0,0)} &= 
    \frac{\left(g_L+g_R\right)e^{2} m_t\cos(\theta)}{2 M_W^{2} s_W^{2}} \mathcal{E}\\
\mathcal{M}_{Z}^{(-\frac{1}{2},\frac{1}{2},-1,0)} &= 
    \frac{g_Le^{2} \left(1-\cos(\theta)\right)}{\sqrt{2}M_Ws_W^{2} }\mathcal{E} \\
\mathcal{M}_{Z}^{(-\frac{1}{2},\frac{1}{2},0,-1)} &= 
    -\frac{g_Le^{2} \left(1+\cos(\theta)\right)}{\sqrt{2}M_Ws_W^{2} }\mathcal{E} \\
\mathcal{M}_{Z}^{(-\frac{1}{2},\frac{1}{2},0,0)} &= 
    \frac{g_Le^{2}\sin(\theta)}{ M_W^{2} s_W^{2}} \mathcal{E}^{2} \\
\mathcal{M}_{Z}^{(-\frac{1}{2},\frac{1}{2},0,1)} &= 
    -\frac{\sqrt{2}g_Le^{2} \sin^{2}\left(\frac{\theta}{2}\right) }{M_Ws_W^{2} } \mathcal{E}\\
\mathcal{M}_{Z}^{(-\frac{1}{2},\frac{1}{2},1,0)} &= 
    \frac{g_Le^{2} \left(1+\cos(\theta)\right)}{\sqrt{2}M_Ws_W^{2} } \mathcal{E}\\
\mathcal{M}_{Z}^{(\frac{1}{2},-\frac{1}{2},-1,0)} &= 
    -\frac{g_Re^{2} \left(1+\cos(\theta)\right)}{\sqrt{2}M_Ws_W^{2} } \mathcal{E}\\
\mathcal{M}_{Z}^{(\frac{1}{2},-\frac{1}{2},0,-1)} &= 
    \frac{g_Re^{2} \left(1-\cos(\theta)\right)}{\sqrt{2}M_Ws_W^{2} } \mathcal{E}\\
\mathcal{M}_{Z}^{(\frac{1}{2},-\frac{1}{2},0,0)} &= 
    \frac{g_Re^{2}\sin(\theta)}{ M_W^{2} s_W^{2}} \mathcal{E}^{2} \\
\mathcal{M}_{Z}^{(\frac{1}{2},-\frac{1}{2},0,1)} &= 
    \frac{g_Re^{2} \left(1+\cos(\theta)\right)}{\sqrt{2}M_Ws_W^{2} } \mathcal{E}\\
\mathcal{M}_{Z}^{(\frac{1}{2},-\frac{1}{2},1,0)} &= 
    -\frac{\sqrt{2}g_Re^{2} \sin^{2}\left(\frac{\theta}{2}\right) }{M_Ws_W^{2} } \mathcal{E}\\
\mathcal{M}_{Z}^{(\frac{1}{2},\frac{1}{2},0,0)} &= 
    -\frac{\left(g_L+g_R\right)e^{2} m_t\cos(\theta)}{2 M_W^{2} s_W^{2}} \mathcal{E} ,
\end{align}
where $g_L=-2Qs_W^2+1$ and $g_R=-2Qs_W^2$ for the top quark.

After combining the contributions, the $(-\frac{1}{2},-\frac{1}{2},0,0)$ and $(\frac{1}{2},\frac{1}{2},0,0)$ polarization combinations are proportional to
\begin{equation}
    \left(4Qs_W^{2} +g_L+g_R-1\right) = 0 .
    \label{eq:ttWW:4Qs^2+gL+gR-1}
\end{equation}
The $(-\frac{1}{2},\frac{1}{2},\pm1,0)$, $(-\frac{1}{2},\frac{1}{2},0,\pm1)$ and $(-\frac{1}{2},\frac{1}{2},0,0)$ polarization combinations are proportional to
\begin{equation}
    \left(2Qs_W^{2}+g_L-1 \right) = 0 .
    \label{eq:ttWW:2Qs^2+gL-1}
\end{equation}
The $(\frac{1}{2},-\frac{1}{2},\pm1,0)$, $(\frac{1}{2},-\frac{1}{2},0,\pm1)$ and $(\frac{1}{2},-\frac{1}{2},0,0)$ polarization combinations are proportional to
\begin{equation}
    \left(2Qs_W^{2}+g_R \right) = 0 .
    \label{eq:ttWW:2Qs^2+gR}
\end{equation}
All the high-energy growth terms cancel.
We did this for the top quark, but a similar relation holds for all the fermions.  

Furthermore, we find agreement at all energies in SPINAS. There is no need or tolerance for a 4-point vertex here.  We have also compared separately the contributions from the internal Higgs, bottom quark, photon, and $Z$ boson and we find agreement in all cases.  In this process, the diagrams are in one-to-one correspondence.

\subsection{$\mathbf{t \bar{t} \to Z Z}$}
The contribution from the Higgs in the S channel is 
\begin{equation}
\mathcal{M}_h =
    -\frac{e^{2} m_t}{2M_W^{2} s_W^{2}} \frac{\langle\mathbf{3}\mathbf{4}\rangle \lbrack\mathbf{3}\mathbf{4}\rbrack
    \left(\langle\mathbf{1}\mathbf{2}\rangle 
    + \lbrack\mathbf{1}\mathbf{2}\rbrack \right)}{ \left(s-m_h^{2}\right)} .
\end{equation}
Taylor expanding gives
\begin{align}
\mathcal{M}_{h}^{(-\frac{1}{2},-\frac{1}{2},0,0)} &= 
    \frac{e^{2} m_t}{2 M_W^{2} s_W^{2}} \mathcal{E}\\
\mathcal{M}_{h}^{(\frac{1}{2},\frac{1}{2},0,0)} &=  
    -\frac{e^{2} m_t}{2 M_W^{2} s_W^{2}} \mathcal{E} .
\end{align}

The contributions from the top quark in the T and U channels are
\begin{align}
\mathcal{M}_{Tt} &=
    -\frac{e^{2} }{2 M_W^{2} s_W^{2}(t-m_t^2)}
    \Big(
    \nonumber\\
    &g_L^{2} \left(M_Z\langle\mathbf{3}\mathbf{4}\rangle -\lbrack\mathbf{3}\lvert p_{1} \rvert \mathbf{4}\rangle \right) \langle\mathbf{1}\mathbf{3}\rangle \lbrack\mathbf{2}\mathbf{4}\rbrack  
    \nonumber\\
    &+g_R^{2} \left(M_Z\lbrack\mathbf{3}\mathbf{4}\rbrack -\lbrack\mathbf{4}\lvert p_{1} \rvert \mathbf{3}\rangle \right)\langle\mathbf{2}\mathbf{4}\rangle \lbrack\mathbf{1}\mathbf{3}\rbrack 
    \label{eq:ttZZ:t T}
    \nonumber\\
    &-g_Lg_Rm_t\left(\langle\mathbf{3}\mathbf{4}\rangle \lbrack\mathbf{1}\mathbf{3}\rbrack \lbrack\mathbf{2}\mathbf{4}\rbrack +\langle\mathbf{1}\mathbf{3}\rangle \langle\mathbf{2}\mathbf{4}\rangle \lbrack\mathbf{3}\mathbf{4}\rbrack \right)
    \Big)
    \\
\mathcal{M}_{Ut} &=
    -\frac{e^{2} }{2 M_W^{2} s_W^{2}(u-m_t^2)}
    \Big(
    \nonumber\\
    &-g_R^{2}\left(M_Z\lbrack\mathbf{3}\mathbf{4}\rbrack +\lbrack\mathbf{3}\lvert p_{1} \rvert \mathbf{4}\rangle \right)  \langle\mathbf{2}\mathbf{3}\rangle \lbrack\mathbf{1}\mathbf{4}\rbrack  
    \nonumber\\
    &-g_L^{2}\left(M_Z\langle\mathbf{3}\mathbf{4}\rangle +\lbrack\mathbf{4}\lvert p_{1} \rvert \mathbf{3}\rangle \right)  \langle\mathbf{1}\mathbf{4}\rangle \lbrack\mathbf{2}\mathbf{3}\rbrack  
    \nonumber\\
    &+g_Lg_Rm_t\left(\langle\mathbf{3}\mathbf{4}\rangle \lbrack\mathbf{1}\mathbf{4}\rbrack \lbrack\mathbf{2}\mathbf{3}\rbrack +\langle\mathbf{1}\mathbf{4}\rangle \langle\mathbf{2}\mathbf{3}\rangle \lbrack\mathbf{3}\mathbf{4}\rbrack \right)
    \Big) .
    \label{eq:ttZZ:t U}
\end{align}
After Taylor expanding in high energy and combining the T and U channels, we have
\begin{align}
\mathcal{M}_{t}^{(-\frac{1}{2},-\frac{1}{2},0,0)} &= 
    -\frac{e^{2} \left(g_L-g_R\right)^{2} m_t}{2 M_W^{2} s_W^{2}} \mathcal{E}\\
\mathcal{M}_{t}^{(\frac{1}{2},\frac{1}{2},0,0)} &= 
    \frac{e^{2} \left(g_L-g_R\right)^{2} m_t}{2 M_W^{2} s_W^{2}} \mathcal{E} .
\end{align}

Combining all the contributions, we see that the high-energy growth term is proportional to $(g_L-g_R)^2-1$, which vanishes.

Moreover, we have compared the constructive amplitude with Feynman diagrams for all energies and found agreement in SPINAS.  No 4-point vertex is needed or allowed here.  We have also tested and find the contribution from an internal Higgs and separately from an internal top quark agree with Feynman diagrams.  No rearrangement of contributions is present here.

\subsection{$\mathbf{\bar{t} b \to Z \bar{W}}$}
The contribution from the $W$ boson in the S channel is
\begin{align}
\mathcal{M}_W &= 
    \frac{-e^2}{\sqrt{2}M_W^{2} M_Z^{2} s_W^{2} \left(s-M_W^{2}\right)}
    \Big(
    \nonumber\\
    &2M_W^{3} 
    \left(
    \langle\mathbf{2}\mathbf{4}\rangle \langle\mathbf{3}\mathbf{4}\rangle \lbrack\mathbf{1}\mathbf{3}\rbrack 
    + \langle\mathbf{2}\mathbf{3}\rangle \lbrack\mathbf{1}\mathbf{4}\rbrack \lbrack\mathbf{3}\mathbf{4}\rbrack
    \right)
    \nonumber\\
    &+2M_W^{2} M_Z \left(
    \langle\mathbf{2}\mathbf{3}\rangle \langle\mathbf{3}\mathbf{4}\rangle \lbrack\mathbf{1}\mathbf{4}\rbrack
    + \langle\mathbf{2}\mathbf{4}\rangle \lbrack\mathbf{1}\mathbf{3}\rbrack \lbrack\mathbf{3}\mathbf{4}\rbrack
    \right)
    \nonumber\\
    &+m_t\left(M_Z^2-2M_W^{2}\right) \langle\mathbf{1}\mathbf{2}\rangle \langle\mathbf{3}\mathbf{4}\rangle \lbrack\mathbf{3}\mathbf{4}\rbrack 
    \nonumber\\
    &-m_b\left(M_Z^2-2M_W^{2}\right) \langle\mathbf{3}\mathbf{4}\rangle \lbrack\mathbf{1}\mathbf{2}\rbrack \lbrack\mathbf{3}\mathbf{4}\rbrack 
    \nonumber\\
    &+2M_W^{2} \langle\mathbf{3}\mathbf{4}\rangle \lbrack\mathbf{3}\mathbf{4}\rbrack \lbrack\mathbf{1}\lvert p_{3} \rvert \mathbf{2}\rangle 
    \Big) .
    \label{eq:tbZW:W S}
\end{align}
Taylor expanding gives us
\begin{align}
\mathcal{M}_{W}^{(-\frac{1}{2},-\frac{1}{2},0,0)} &= 
    \frac{e^2m_t\left(1-c_W^{2} +c_W^{2} \cos(\theta)\right)}{\sqrt{2} c_W^{2}  M_Z^{2} s_W^{2}} \mathcal{E}\\
\mathcal{M}_{W}^{(\frac{1}{2},-\frac{1}{2},-1,0)} &= 
    -\frac{e^{2} \left(1+\cos(\theta)\right)\mathcal{E}}{M_Zs_W^{2} } \\
\mathcal{M}_{W}^{(\frac{1}{2},-\frac{1}{2},0,-1)} &= 
    \frac{e^{2}c_W \left(1-\cos(\theta)\right)\mathcal{E}}{M_Zs_W^{2} } \\
\mathcal{M}_{W}^{(\frac{1}{2},-\frac{1}{2},0,0)} &= 
    \frac{\sqrt{2}e^{2}\sin(\theta)}{ M_Z^{2} s_W^{2}} \mathcal{E}^{2} \\
\mathcal{M}_{W}^{(\frac{1}{2},-\frac{1}{2},0,1)} &= 
    \frac{e^{2}c_W \left(1+\cos(\theta)\right)\mathcal{E}}{M_Zs_W^{2} } \\
\mathcal{M}_{W}^{(\frac{1}{2},-\frac{1}{2},1,0)} &= 
    \frac{e^{2} \left(-1+\cos(\theta)\right)\mathcal{E}}{M_Zs_W^{2} } \\
\mathcal{M}_{W}^{(\frac{1}{2},\frac{1}{2},0,0)} &= 
    \frac{e^{2} m_b\left(1-c_W^{2} -c_W^{2} \cos(\theta)\right)}{\sqrt{2} c_W^{2} M_Z^{2} s_W^{2}} \mathcal{E} .
\end{align}

The contribution from a top quark in the T channel is
\begin{align}
\mathcal{M}_t &= 
    \frac{e^2\langle\mathbf{2}\mathbf{4}\rangle}{\sqrt{2}M_W^{2} s_W^{2}\left(t-m_t^{2}\right)}
    \Big(g_{Ru}m_t\langle\mathbf{1}\mathbf{3}\rangle \lbrack\mathbf{3}\mathbf{4}\rbrack 
    \nonumber\\
    &-g_{Lu}\lbrack\mathbf{1}\mathbf{3}\rbrack\left(M_Z\lbrack\mathbf{3}\mathbf{4}\rbrack -\lbrack\mathbf{4}\lvert p_{1} \rvert \mathbf{3}\rangle \right) \Big) .
    \label{eq:tbZW:t T}
\end{align}
After Taylor expanding,
\begin{align}
\mathcal{M}_{t}^{(-\frac{1}{2},-\frac{1}{2},0,0)} &= 
	-\frac{e^{2} m_t\left(g_{Lu}-2g_{Ru}+g_{Lu}\cos(\theta)\right)}{2\sqrt{2} c_W^{2} M_Z^{2} s_W^{2}} \mathcal{E}\\
\mathcal{M}_{t}^{(\frac{1}{2},-\frac{1}{2},-1,0)} &= 
	\frac{e^{2} g_{Lu}\left(1+\cos(\theta)\right)}{2c_W^{2} M_Zs_W^{2} } \mathcal{E}\\
\mathcal{M}_{t}^{(\frac{1}{2},-\frac{1}{2},0,-1)} &= 
	-\frac{e^{2} g_{Lu}\sin^{2}\left(\frac{\theta}{2}\right) }{c_WM_Zs_W^{2} }\mathcal{E} \\
\mathcal{M}_{t}^{(\frac{1}{2},-\frac{1}{2},0,0)} &= 
	-\frac{e^{2} g_{Lu}\sin(\theta)}
    {\sqrt{2} c_W^{2} M_Z^{2} s_W^{2}} \mathcal{E}^{2} \\
\mathcal{M}_{t}^{(\frac{1}{2},-\frac{1}{2},0,1)} &= 
	-\frac{e^{2} g_{Lu}\left(1+\cos(\theta)\right)}{2c_WM_Zs_W^{2} }\mathcal{E} \\
\mathcal{M}_{t}^{(\frac{1}{2},-\frac{1}{2},1,0)} &= 
	\frac{e^{2} g_{Lu}\left(1-\cos(\theta)\right)}{2c_W^{2} M_Zs_W^{2} } \mathcal{E}\\
\mathcal{M}_{t}^{(\frac{1}{2},\frac{1}{2},0,0)} &= 
	\frac{e^{2} g_{Lu}m_b\left(1+\cos(\theta)\right)}{2\sqrt{2} c_W^{2} M_Z^{2} s_W^{2}} \mathcal{E} .
\end{align}

The contribution from the $b$ quark in the U channel is
\begin{align}
\mathcal{M}_b &= 
    \frac{-e^2\lbrack\mathbf{1}\mathbf{4}\rbrack}{\sqrt{2}M_W^{2} s_W^{2}(u-m_b^2)}
     \Big(
     g_{Rd}m_b\langle\mathbf{3}\mathbf{4}\rangle \lbrack\mathbf{2}\mathbf{3}\rbrack 
     \nonumber\\
     &-g_{Ld}\langle\mathbf{2}\mathbf{3}\rangle\left(M_W\lbrack\mathbf{3}\mathbf{4}\rbrack +\lbrack\mathbf{3}\lvert p_{1} \rvert \mathbf{4}\rangle \right) \Big) .
    \label{eq:tbZW:b U}
\end{align}
Taylor expanding, we have
\begin{align}
\mathcal{M}_{b}^{(-\frac{1}{2},-\frac{1}{2},0,0)} &= 
    -\frac{e^{2} g_{Ld}m_t\sin^{2}\left(\frac{\theta}{2}\right)}
    {\sqrt{2} c_W^{2} M_Z^{2} s_W^{2}} \mathcal{E} \\
\mathcal{M}_{b}^{(\frac{1}{2},-\frac{1}{2},-1,0)} &= 
	-\frac{e^{2} g_{Ld}\left(1+\cos(\theta)\right)}{2c_W^{2} M_Zs_W^{2} } \mathcal{E}\\
\mathcal{M}_{b}^{(\frac{1}{2},-\frac{1}{2},0,-1)} &= 
	\frac{e^{2} g_{Ld}\left(1-\cos(\theta)\right)}{2c_WM_Zs_W^{2} } \mathcal{E}\\
\mathcal{M}_{b}^{(\frac{1}{2},-\frac{1}{2},0,0)} &= 
	\frac{e^{2} g_{Ld}\sin(\theta)}{\sqrt{2} c_W^{2} M_Z^{2} s_W^{2}} \mathcal{E}^{2} \\
\mathcal{M}_{b}^{(\frac{1}{2},-\frac{1}{2},0,1)} &= 
	\frac{e^{2} g_{Ld}\left(1+\cos(\theta)\right)}{2c_WM_Zs_W^{2} }\mathcal{E} \\
\mathcal{M}_{b}^{(\frac{1}{2},-\frac{1}{2},1,0)} &= 
	-\frac{e^{2} g_{Ld}\sin^{2}\left(\frac{\theta}{2}\right) }{c_W^{2} M_Zs_W^{2} }\mathcal{E} \\
\mathcal{M}_{b}^{(\frac{1}{2},\frac{1}{2},0,0)} &= 
	-\frac{e^{2} m_b\left(2g_{Rd}-g_{Ld}+g_{Ld}\cos(\theta)\right)}{2\sqrt{2} c_W^{2} M_Z^{2} s_W^{2} }\mathcal{E} .
\end{align}

After combining, every term is proportional to one of
\begin{align}
    2 c_W^2+g_{Ld}-g_{Lu} &= 
    2 c_W^2+2(Q_u-Q_d)s_W^2-2\\
    2 c_W^2+g_{Ld}+g_{Lu}-2 g_{Ru} -2 &= 
    2 c_W^2+2(Q_u-Q_d)s_W^2 -2\\
    2 c_W^2-g_{Ld}-g_{Lu}+2 g_{Rd} - 2 &= 
    2 c_W^2+2(Q_u-Q_d)s_W^2 - 2
\end{align}
and they all vanish.  We have also validated this process at all energies and find agreement with Feynman diagrams.  No 4-point vertex is needed or allowed.  Further, we have tested the diagrams individually and they agree with Feynman diagrams, so there is no rearrangement of contributions here.

\section{\label{sec:conclusions}Conclusions}
In this paper, we have analyzed perturbative unitarity in all the processes involving a longitudinal vector boson.  We have found the high-energy expansion of all the CSM contributions coming from S-, T- and U-channel diagrams using constructive techniques.  In the cases that these do not cancel on their own, we have identified the 4-point vertex required to cancel them.  Ultimately, we have shown that perturbative unitarity is satisfied in all these processes and we have found a complete set of 4-point vertices in the CSM (shown in Table~\ref{tab:4-point vertices}) in order to achieve these cancellations.  We have included the 4-point vertex for four Higgs bosons for completeness.  

We have shown that, not only do we not need more 4-point vertices in the CSM relative to Feynman diagrams, we actually need fewer.  It was already well known that the CSM does not require a 4-point gluon vertex, but we additionally show in this paper that we also do not need a $Z Z \bar{W} W$ vertex, a $W W \bar{W} \bar{W}$ vertex, a $\gamma Z \bar{W} W$ vertex or a $\gamma \gamma \bar{W} W$ vertex.  In other words, we explicitly find that we do not need any ``contact" terms at four points, especially if we note that we did not need any other 4-point vertices (or contact terms) for any other 4-point amplitudes in the CSM that were given in \cite{Christensen:2024C}.  In fact, we claim that this is what we should expect in a renormalizable theory like the SM, even when constructive, and we do not expect to need any extra contact terms at any order in external particles or loops in the CSM.  If we did need further 4-point vertices (or contact terms) at higher loop level, then those 4-point vertices would arguably ruin the tree-level 4-point amplitudes that don't allow them.  Showing this is a goal of future research.

We have further shown that there is a significant rearrangement of contributions to the amplitude in many cases, especially those involving a 4-point vertex in Feynman diagrams but do not have a 4-point vertex in constructive calculations.  In particular, we see that the constructive calculations for $Z Z \to W \bar{W}$, $W W \to W W$, $\gamma Z W \bar{W}$ and $\gamma \gamma W \bar{W}$ do not even allow a 4-point vertex in the CSM, but it is required for Feynman diagrams.  This shows that constructive diagrams are not in one-to-one correspondence with Feynman diagrams in general, even when there is a superficial resemblance.  In many cases, the contributions to the complete amplitude are rearranged.   There was already a hint that this was true with processes with external photons or gluons and we have found agreement with this more generally, but we have additionally shown that it is also true for some processes that do not include external photons or gluons.  In particular, we have shown it for amplitudes with four external $Z$ or $W$ bosons.  We expect this to compound as the number of external number of particles grows, and understanding this will be one of the goals of future projects.  We have shown it explicitly in their high-energy expansions here.  But, we have also found it for all energies using SPINAS.  

Finally, we have noted that the cancellations are under much better control in these constructive calculations compared to Feynman diagrams.  First, there is no quartic energy growth at all in the constructive amplitudes for $Z Z \to W \bar{W}$ and $W W \to W W$, unlike in Feynman diagrams.  Further, we have found that the cancellations that do take place are mostly between pairs of diagrams involving the same internal particle, whereas in Feynman diagrams, a larger set of particles involving a 4-point vertex and different internal particles are typically involved.  The final cancellation between diagrams with different particles is only quadratic and only in the all-longitudinal channel.  Furthermore, if the process involves an external photon or gluon, there is no high-energy growth at all.  Therefore, there is no cancellation at all.  In fact, there is only one term in the expression with both propagator denominators and every term does not have energy growth.  All of this should lead to better numerical stability in phase-space integrators.

\appendix

\begin{widetext}
\section{\label{app:amplitudes}Long Contributions to $\mathbf{Z Z \to W \bar{W}}$ and $\mathbf{W \bar{W} \to W \bar{W}}$}
In this appendix, we will give the full expressions for the longer contributions to $Z Z \to W \bar{W}$ and $W \bar{W} \to W \bar{W}$.  The expressions when the triple-boson vertices are involved are quite lengthy, so we have included them here to improve the readability of the main text.

\subsection{\label{app:ZZWW 4: Derivation}Long Contribution to $\mathbf{Z Z \to W \bar{W}}$}

The contributions to the amplitude for $Z Z \to W \bar{W}$ coming from a T- and a U-channel $W$ boson are 
\begin{align}
\mathcal{M}_{TW} &= \frac{e^2}{2M_W^3s_W^2\left(t-M_W^2\right)}
\Bigg(
    \Big(
        -4c_W^{3} \langle\mathbf{2}\mathbf{4}\rangle \langle\mathbf{3}\mathbf{4}\rangle \lbrack\mathbf{1}\mathbf{2}\rbrack \lbrack\mathbf{1}\mathbf{3}\rbrack 
        +3c_W^{3} \langle\mathbf{1}\mathbf{3}\rangle \langle\mathbf{2}\mathbf{3}\rangle \langle\mathbf{2}\mathbf{4}\rangle \lbrack\mathbf{1}\mathbf{4}\rbrack 
        -4c_W^{2} \langle\mathbf{2}\mathbf{3}\rangle \langle\mathbf{2}\mathbf{4}\rangle \lbrack\mathbf{1}\mathbf{3}\rbrack \lbrack\mathbf{1}\mathbf{4}\rbrack 
        \nonumber\\
        &-3c_W^{4} \langle\mathbf{1}\mathbf{4}\rangle \langle\mathbf{2}\mathbf{4}\rangle \lbrack\mathbf{1}\mathbf{3}\rbrack \lbrack\mathbf{2}\mathbf{3}\rbrack 
        -3c_W^{2} \langle\mathbf{1}\mathbf{3}\rangle \langle\mathbf{2}\mathbf{4}\rangle \lbrack\mathbf{1}\mathbf{4}\rbrack \lbrack\mathbf{2}\mathbf{3}\rbrack 
        -c_W^{4} \langle\mathbf{1}\mathbf{3}\rangle \langle\mathbf{2}\mathbf{4}\rangle \lbrack\mathbf{1}\mathbf{4}\rbrack \lbrack\mathbf{2}\mathbf{3}\rbrack 
        +c_W^{3} \langle\mathbf{2}\mathbf{4}\rangle \lbrack\mathbf{1}\mathbf{3}\rbrack \lbrack\mathbf{1}\mathbf{4}\rbrack \lbrack\mathbf{2}\mathbf{3}\rbrack 
        \nonumber\\
        &+3c_W^{3} \langle\mathbf{1}\mathbf{3}\rangle \langle\mathbf{1}\mathbf{4}\rangle \langle\mathbf{2}\mathbf{3}\rangle \lbrack\mathbf{2}\mathbf{4}\rbrack 
        -3c_W^{2} \langle\mathbf{1}\mathbf{4}\rangle \langle\mathbf{2}\mathbf{3}\rangle \lbrack\mathbf{1}\mathbf{3}\rbrack \lbrack\mathbf{2}\mathbf{4}\rbrack 
        -c_W^{4} \langle\mathbf{1}\mathbf{4}\rangle \langle\mathbf{2}\mathbf{3}\rangle \lbrack\mathbf{1}\mathbf{3}\rbrack \lbrack\mathbf{2}\mathbf{4}\rbrack 
        -2\langle\mathbf{1}\mathbf{3}\rangle \langle\mathbf{2}\mathbf{4}\rangle \lbrack\mathbf{1}\mathbf{3}\rbrack \lbrack\mathbf{2}\mathbf{4}\rbrack 
        \nonumber\\
        &+6c_W^{2} \langle\mathbf{1}\mathbf{3}\rangle \langle\mathbf{2}\mathbf{4}\rangle \lbrack\mathbf{1}\mathbf{3}\rbrack \lbrack\mathbf{2}\mathbf{4}\rbrack 
        -2c_W^{4} \langle\mathbf{1}\mathbf{3}\rangle \langle\mathbf{2}\mathbf{4}\rangle \lbrack\mathbf{1}\mathbf{3}\rbrack \lbrack\mathbf{2}\mathbf{4}\rbrack 
        -2c_W^{2} \langle\mathbf{1}\mathbf{2}\rangle \langle\mathbf{3}\mathbf{4}\rangle \lbrack\mathbf{1}\mathbf{3}\rbrack \lbrack\mathbf{2}\mathbf{4}\rbrack 
        +2c_W^{4} \langle\mathbf{1}\mathbf{2}\rangle \langle\mathbf{3}\mathbf{4}\rangle \lbrack\mathbf{1}\mathbf{3}\rbrack \lbrack\mathbf{2}\mathbf{4}\rbrack 
        \nonumber\\
        &-c_W^{4} \langle\mathbf{1}\mathbf{3}\rangle \langle\mathbf{2}\mathbf{3}\rangle \lbrack\mathbf{1}\mathbf{4}\rbrack \lbrack\mathbf{2}\mathbf{4}\rbrack 
        -4c_W^{2} \langle\mathbf{1}\mathbf{3}\rangle \langle\mathbf{1}\mathbf{4}\rangle \lbrack\mathbf{2}\mathbf{3}\rbrack \lbrack\mathbf{2}\mathbf{4}\rbrack 
        +c_W^{3} \langle\mathbf{1}\mathbf{4}\rangle \lbrack\mathbf{1}\mathbf{3}\rbrack \lbrack\mathbf{2}\mathbf{3}\rbrack \lbrack\mathbf{2}\mathbf{4}\rbrack 
        -4c_W^{3} \langle\mathbf{1}\mathbf{2}\rangle \langle\mathbf{1}\mathbf{3}\rangle \lbrack\mathbf{2}\mathbf{4}\rbrack \lbrack\mathbf{3}\mathbf{4}\rbrack 
    \Big)M_W
    \nonumber\\
    &+\Big(
        -3c_W^{4} \langle\mathbf{1}\mathbf{3}\rangle \langle\mathbf{2}\mathbf{4}\rangle \lbrack\mathbf{2}\mathbf{3}\rbrack 
        -3c_W^{4} \langle\mathbf{1}\mathbf{3}\rangle \langle\mathbf{2}\mathbf{3}\rangle \lbrack\mathbf{2}\mathbf{4}\rbrack 
        +c_W^{3} \langle\mathbf{2}\mathbf{3}\rangle \lbrack\mathbf{1}\mathbf{3}\rbrack \lbrack\mathbf{2}\mathbf{4}\rbrack 
        +c_W^{3} \langle\mathbf{1}\mathbf{3}\rangle \lbrack\mathbf{2}\mathbf{3}\rbrack \lbrack\mathbf{2}\mathbf{4}\rbrack 
    \Big)\lbrack\mathbf{1}\lvert p_{3} \rvert \mathbf{4}\rangle 
    \nonumber\\
    &+\Big(
        3c_W^{3} \langle\mathbf{2}\mathbf{3}\rangle \langle\mathbf{2}\mathbf{4}\rangle \lbrack\mathbf{1}\mathbf{3}\rbrack 
        +3c_W^{3} \langle\mathbf{1}\mathbf{3}\rangle \langle\mathbf{2}\mathbf{4}\rangle \lbrack\mathbf{2}\mathbf{3}\rbrack 
        -c_W^{4} \langle\mathbf{2}\mathbf{4}\rangle \lbrack\mathbf{1}\mathbf{3}\rbrack \lbrack\mathbf{2}\mathbf{3}\rbrack 
        -c_W^{4} \langle\mathbf{2}\mathbf{3}\rangle \lbrack\mathbf{1}\mathbf{3}\rbrack \lbrack\mathbf{2}\mathbf{4}\rbrack 
    \Big)\lbrack\mathbf{4}\lvert p_{3} \rvert \mathbf{1}\rangle
\Bigg)
\label{eq:app:4-point:ZZWW:WT}
\end{align}

\begin{align}
\mathcal{M}_{UW} &= \frac{e^2}{2M_W^3s_W^2\left(u-M_W^2\right)}
\Bigg(
    \Big(
        3c_W^{3} \langle\mathbf{1}\mathbf{4}\rangle \langle\mathbf{2}\mathbf{3}\rangle \langle\mathbf{2}\mathbf{4}\rangle \lbrack\mathbf{1}\mathbf{3}\rbrack 
        +4c_W^{3} \langle\mathbf{2}\mathbf{3}\rangle \langle\mathbf{3}\mathbf{4}\rangle \lbrack\mathbf{1}\mathbf{2}\rbrack \lbrack\mathbf{1}\mathbf{4}\rbrack 
        -4c_W^{2} \langle\mathbf{2}\mathbf{3}\rangle \langle\mathbf{2}\mathbf{4}\rangle \lbrack\mathbf{1}\mathbf{3}\rbrack \lbrack\mathbf{1}\mathbf{4}\rbrack 
        \nonumber\\
        &+3c_W^{3} \langle\mathbf{1}\mathbf{3}\rangle \langle\mathbf{1}\mathbf{4}\rangle \langle\mathbf{2}\mathbf{4}\rangle \lbrack\mathbf{2}\mathbf{3}\rbrack 
        -c_W^{4} \langle\mathbf{1}\mathbf{4}\rangle \langle\mathbf{2}\mathbf{4}\rangle \lbrack\mathbf{1}\mathbf{3}\rbrack \lbrack\mathbf{2}\mathbf{3}\rbrack 
        -2\langle\mathbf{1}\mathbf{4}\rangle \langle\mathbf{2}\mathbf{3}\rangle \lbrack\mathbf{1}\mathbf{4}\rbrack \lbrack\mathbf{2}\mathbf{3}\rbrack 
        +6c_W^{2} \langle\mathbf{1}\mathbf{4}\rangle \langle\mathbf{2}\mathbf{3}\rangle \lbrack\mathbf{1}\mathbf{4}\rbrack \lbrack\mathbf{2}\mathbf{3}\rbrack 
        \nonumber\\
        &-2c_W^{4} \langle\mathbf{1}\mathbf{4}\rangle \langle\mathbf{2}\mathbf{3}\rangle \lbrack\mathbf{1}\mathbf{4}\rbrack \lbrack\mathbf{2}\mathbf{3}\rbrack 
        -3c_W^{2} \langle\mathbf{1}\mathbf{3}\rangle \langle\mathbf{2}\mathbf{4}\rangle \lbrack\mathbf{1}\mathbf{4}\rbrack \lbrack\mathbf{2}\mathbf{3}\rbrack 
        -c_W^{4} \langle\mathbf{1}\mathbf{3}\rangle \langle\mathbf{2}\mathbf{4}\rangle \lbrack\mathbf{1}\mathbf{4}\rbrack \lbrack\mathbf{2}\mathbf{3}\rbrack 
        +2c_W^{2} \langle\mathbf{1}\mathbf{2}\rangle \langle\mathbf{3}\mathbf{4}\rangle \lbrack\mathbf{1}\mathbf{4}\rbrack \lbrack\mathbf{2}\mathbf{3}\rbrack 
        \nonumber\\
        &-2c_W^{4} \langle\mathbf{1}\mathbf{2}\rangle \langle\mathbf{3}\mathbf{4}\rangle \lbrack\mathbf{1}\mathbf{4}\rbrack \lbrack\mathbf{2}\mathbf{3}\rbrack 
        -3c_W^{2} \langle\mathbf{1}\mathbf{4}\rangle \langle\mathbf{2}\mathbf{3}\rangle \lbrack\mathbf{1}\mathbf{3}\rbrack \lbrack\mathbf{2}\mathbf{4}\rbrack 
        -c_W^{4} \langle\mathbf{1}\mathbf{4}\rangle \langle\mathbf{2}\mathbf{3}\rangle \lbrack\mathbf{1}\mathbf{3}\rbrack \lbrack\mathbf{2}\mathbf{4}\rbrack 
        -3c_W^{4} \langle\mathbf{1}\mathbf{3}\rangle \langle\mathbf{2}\mathbf{3}\rangle \lbrack\mathbf{1}\mathbf{4}\rbrack \lbrack\mathbf{2}\mathbf{4}\rbrack 
        \nonumber\\
        &+c_W^{3} \langle\mathbf{2}\mathbf{3}\rangle \lbrack\mathbf{1}\mathbf{3}\rbrack \lbrack\mathbf{1}\mathbf{4}\rbrack \lbrack\mathbf{2}\mathbf{4}\rbrack 
        -4c_W^{2} \langle\mathbf{1}\mathbf{3}\rangle \langle\mathbf{1}\mathbf{4}\rangle \lbrack\mathbf{2}\mathbf{3}\rbrack \lbrack\mathbf{2}\mathbf{4}\rbrack 
        +c_W^{3} \langle\mathbf{1}\mathbf{3}\rangle \lbrack\mathbf{1}\mathbf{4}\rbrack \lbrack\mathbf{2}\mathbf{3}\rbrack \lbrack\mathbf{2}\mathbf{4}\rbrack 
        +4c_W^{3} \langle\mathbf{1}\mathbf{2}\rangle \langle\mathbf{1}\mathbf{4}\rangle \lbrack\mathbf{2}\mathbf{3}\rbrack \lbrack\mathbf{3}\mathbf{4}\rbrack 
    \Big)M_W
    \nonumber\\
    &+\Big(
        -3c_W^{4} \langle\mathbf{1}\mathbf{4}\rangle \langle\mathbf{2}\mathbf{4}\rangle \lbrack\mathbf{2}\mathbf{3}\rbrack 
        +c_W^{3} \langle\mathbf{2}\mathbf{4}\rangle \lbrack\mathbf{1}\mathbf{4}\rbrack \lbrack\mathbf{2}\mathbf{3}\rbrack 
        -3c_W^{4} \langle\mathbf{1}\mathbf{4}\rangle \langle\mathbf{2}\mathbf{3}\rangle \lbrack\mathbf{2}\mathbf{4}\rbrack 
        +c_W^{3} \langle\mathbf{1}\mathbf{4}\rangle \lbrack\mathbf{2}\mathbf{3}\rbrack \lbrack\mathbf{2}\mathbf{4}\rbrack 
    \Big)\lbrack\mathbf{1}\lvert p_{4} \rvert \mathbf{3}\rangle 
    \nonumber\\
    &+\Big(
        3c_W^{3} \langle\mathbf{2}\mathbf{3}\rangle \langle\mathbf{2}\mathbf{4}\rangle \lbrack\mathbf{1}\mathbf{4}\rbrack 
        -1c_W^{4} \langle\mathbf{2}\mathbf{4}\rangle \lbrack\mathbf{1}\mathbf{4}\rbrack \lbrack\mathbf{2}\mathbf{3}\rbrack 
        +3c_W^{3} \langle\mathbf{1}\mathbf{4}\rangle \langle\mathbf{2}\mathbf{3}\rangle \lbrack\mathbf{2}\mathbf{4}\rbrack 
        -c_W^{4} \langle\mathbf{2}\mathbf{3}\rangle \lbrack\mathbf{1}\mathbf{4}\rbrack \lbrack\mathbf{2}\mathbf{4}\rbrack 
    \Big)\lbrack\mathbf{3}\lvert p_{4} \rvert \mathbf{1}\rangle
\Bigg) .
\label{eq:app:4-point:ZZWW:WU}
\end{align}

\subsection{Long Contributions to $W \bar{W} \to W \bar{W}$}

The contribution coming from the photon in the T and U channels are 
\begin{align}
\mathcal{M}_{T\gamma} &= \frac{e^2}{M_W^3 t}
\Bigg(
    \Big(
        2\langle\mathbf{2}\mathbf{4}\rangle \langle\mathbf{3}\mathbf{4}\rangle \lbrack\mathbf{1}\mathbf{2}\rbrack \lbrack\mathbf{1}\mathbf{3}\rbrack 
        -\langle\mathbf{1}\mathbf{3}\rangle \langle\mathbf{2}\mathbf{3}\rangle \langle\mathbf{2}\mathbf{4}\rangle \lbrack\mathbf{1}\mathbf{4}\rbrack 
        +2\langle\mathbf{2}\mathbf{3}\rangle \langle\mathbf{2}\mathbf{4}\rangle \lbrack\mathbf{1}\mathbf{3}\rbrack \lbrack\mathbf{1}\mathbf{4}\rbrack 
        +\langle\mathbf{1}\mathbf{4}\rangle \langle\mathbf{2}\mathbf{4}\rangle \lbrack\mathbf{1}\mathbf{3}\rbrack \lbrack\mathbf{2}\mathbf{3}\rbrack 
        +2\langle\mathbf{1}\mathbf{3}\rangle \langle\mathbf{2}\mathbf{4}\rangle \lbrack\mathbf{1}\mathbf{4}\rbrack \lbrack\mathbf{2}\mathbf{3}\rbrack 
        \nonumber\\
        &-\langle\mathbf{2}\mathbf{4}\rangle \lbrack\mathbf{1}\mathbf{3}\rbrack \lbrack\mathbf{1}\mathbf{4}\rbrack \lbrack\mathbf{2}\mathbf{3}\rbrack 
        -\langle\mathbf{1}\mathbf{3}\rangle \langle\mathbf{1}\mathbf{4}\rangle \langle\mathbf{2}\mathbf{3}\rangle \lbrack\mathbf{2}\mathbf{4}\rbrack 
        +2\langle\mathbf{1}\mathbf{4}\rangle \langle\mathbf{2}\mathbf{3}\rangle \lbrack\mathbf{1}\mathbf{3}\rbrack \lbrack\mathbf{2}\mathbf{4}\rbrack 
        -2\langle\mathbf{1}\mathbf{3}\rangle \langle\mathbf{2}\mathbf{4}\rangle \lbrack\mathbf{1}\mathbf{3}\rbrack \lbrack\mathbf{2}\mathbf{4}\rbrack 
        +\langle\mathbf{1}\mathbf{2}\rangle \langle\mathbf{3}\mathbf{4}\rangle \lbrack\mathbf{1}\mathbf{3}\rbrack \lbrack\mathbf{2}\mathbf{4}\rbrack 
        \nonumber\\
        &+\langle\mathbf{1}\mathbf{3}\rangle \langle\mathbf{2}\mathbf{3}\rangle \lbrack\mathbf{1}\mathbf{4}\rbrack \lbrack\mathbf{2}\mathbf{4}\rbrack 
        +2\langle\mathbf{1}\mathbf{3}\rangle \langle\mathbf{1}\mathbf{4}\rangle \lbrack\mathbf{2}\mathbf{3}\rbrack \lbrack\mathbf{2}\mathbf{4}\rbrack 
        -\langle\mathbf{1}\mathbf{4}\rangle \lbrack\mathbf{1}\mathbf{3}\rbrack \lbrack\mathbf{2}\mathbf{3}\rbrack \lbrack\mathbf{2}\mathbf{4}\rbrack 
        +\langle\mathbf{1}\mathbf{3}\rangle \langle\mathbf{2}\mathbf{4}\rangle \lbrack\mathbf{1}\mathbf{2}\rbrack \lbrack\mathbf{3}\mathbf{4}\rbrack 
        +2\langle\mathbf{1}\mathbf{2}\rangle \langle\mathbf{1}\mathbf{3}\rangle \lbrack\mathbf{2}\mathbf{4}\rbrack \lbrack\mathbf{3}\mathbf{4}\rbrack 
    \Big)M_W
    \nonumber\\
    &+\Big(
        \langle\mathbf{1}\mathbf{3}\rangle \langle\mathbf{2}\mathbf{4}\rangle \lbrack\mathbf{2}\mathbf{3}\rbrack 
        +\langle\mathbf{1}\mathbf{3}\rangle \langle\mathbf{2}\mathbf{3}\rangle \lbrack\mathbf{2}\mathbf{4}\rbrack 
        -\langle\mathbf{2}\mathbf{3}\rangle \lbrack\mathbf{1}\mathbf{3}\rbrack \lbrack\mathbf{2}\mathbf{4}\rbrack 
        -\langle\mathbf{1}\mathbf{3}\rangle \lbrack\mathbf{2}\mathbf{3}\rbrack \lbrack\mathbf{2}\mathbf{4}\rbrack 
    \Big)\lbrack\mathbf{1}\lvert p_{3} \rvert \mathbf{4}\rangle 
    \nonumber\\
    &+\Big(
        -\langle\mathbf{2}\mathbf{3}\rangle \langle\mathbf{2}\mathbf{4}\rangle \lbrack\mathbf{1}\mathbf{3}\rbrack 
        -\langle\mathbf{1}\mathbf{3}\rangle \langle\mathbf{2}\mathbf{4}\rangle \lbrack\mathbf{2}\mathbf{3}\rbrack 
        +\langle\mathbf{2}\mathbf{4}\rangle \lbrack\mathbf{1}\mathbf{3}\rbrack \lbrack\mathbf{2}\mathbf{3}\rbrack 
        +\langle\mathbf{2}\mathbf{3}\rangle \lbrack\mathbf{1}\mathbf{3}\rbrack \lbrack\mathbf{2}\mathbf{4}\rbrack 
    \Big)\lbrack\mathbf{4}\lvert p_{3} \rvert \mathbf{1}\rangle
\Bigg)
    \label{eq:WWWW:TA full}
\end{align}
and 
\begin{align}
\mathcal{M}_{U\gamma} &= \frac{e^2}{M_W^3 u}
\Bigg(
    \Big(
        -\langle\mathbf{1}\mathbf{4}\rangle \langle\mathbf{2}\mathbf{3}\rangle \langle\mathbf{2}\mathbf{4}\rangle \lbrack\mathbf{1}\mathbf{3}\rbrack 
        -2\langle\mathbf{2}\mathbf{3}\rangle \langle\mathbf{3}\mathbf{4}\rangle \lbrack\mathbf{1}\mathbf{2}\rbrack \lbrack\mathbf{1}\mathbf{4}\rbrack 
        +2\langle\mathbf{2}\mathbf{3}\rangle \langle\mathbf{2}\mathbf{4}\rangle \lbrack\mathbf{1}\mathbf{3}\rbrack \lbrack\mathbf{1}\mathbf{4}\rbrack 
        -\langle\mathbf{1}\mathbf{3}\rangle \langle\mathbf{1}\mathbf{4}\rangle \langle\mathbf{2}\mathbf{4}\rangle \lbrack\mathbf{2}\mathbf{3}\rbrack 
        +\langle\mathbf{1}\mathbf{4}\rangle \langle\mathbf{2}\mathbf{4}\rangle \lbrack\mathbf{1}\mathbf{3}\rbrack \lbrack\mathbf{2}\mathbf{3}\rbrack 
        \nonumber\\
        &-2\langle\mathbf{1}\mathbf{4}\rangle \langle\mathbf{2}\mathbf{3}\rangle \lbrack\mathbf{1}\mathbf{4}\rbrack \lbrack\mathbf{2}\mathbf{3}\rbrack 
        +2\langle\mathbf{1}\mathbf{3}\rangle \langle\mathbf{2}\mathbf{4}\rangle \lbrack\mathbf{1}\mathbf{4}\rbrack \lbrack\mathbf{2}\mathbf{3}\rbrack 
        -\langle\mathbf{1}\mathbf{2}\rangle \langle\mathbf{3}\mathbf{4}\rangle \lbrack\mathbf{1}\mathbf{4}\rbrack \lbrack\mathbf{2}\mathbf{3}\rbrack 
        +2\langle\mathbf{1}\mathbf{4}\rangle \langle\mathbf{2}\mathbf{3}\rangle \lbrack\mathbf{1}\mathbf{3}\rbrack \lbrack\mathbf{2}\mathbf{4}\rbrack 
        +\langle\mathbf{1}\mathbf{3}\rangle \langle\mathbf{2}\mathbf{3}\rangle \lbrack\mathbf{1}\mathbf{4}\rbrack \lbrack\mathbf{2}\mathbf{4}\rbrack 
        \nonumber\\
        &-\langle\mathbf{2}\mathbf{3}\rangle \lbrack\mathbf{1}\mathbf{3}\rbrack \lbrack\mathbf{1}\mathbf{4}\rbrack \lbrack\mathbf{2}\mathbf{4}\rbrack 
        +2\langle\mathbf{1}\mathbf{3}\rangle \langle\mathbf{1}\mathbf{4}\rangle \lbrack\mathbf{2}\mathbf{3}\rbrack \lbrack\mathbf{2}\mathbf{4}\rbrack 
        -\langle\mathbf{1}\mathbf{3}\rangle \lbrack\mathbf{1}\mathbf{4}\rbrack \lbrack\mathbf{2}\mathbf{3}\rbrack \lbrack\mathbf{2}\mathbf{4}\rbrack 
        -\langle\mathbf{1}\mathbf{4}\rangle \langle\mathbf{2}\mathbf{3}\rangle \lbrack\mathbf{1}\mathbf{2}\rbrack \lbrack\mathbf{3}\mathbf{4}\rbrack 
        -2\langle\mathbf{1}\mathbf{2}\rangle \langle\mathbf{1}\mathbf{4}\rangle \lbrack\mathbf{2}\mathbf{3}\rbrack \lbrack\mathbf{3}\mathbf{4}\rbrack 
    \Big)M_W
    \nonumber\\
    &+\Big(
        \langle\mathbf{1}\mathbf{4}\rangle \langle\mathbf{2}\mathbf{4}\rangle \lbrack\mathbf{2}\mathbf{3}\rbrack 
        -\langle\mathbf{2}\mathbf{4}\rangle \lbrack\mathbf{1}\mathbf{4}\rbrack \lbrack\mathbf{2}\mathbf{3}\rbrack 
        +\langle\mathbf{1}\mathbf{4}\rangle \langle\mathbf{2}\mathbf{3}\rangle \lbrack\mathbf{2}\mathbf{4}\rbrack 
        -\langle\mathbf{1}\mathbf{4}\rangle \lbrack\mathbf{2}\mathbf{3}\rbrack \lbrack\mathbf{2}\mathbf{4}\rbrack 
    \Big)\lbrack\mathbf{1}\lvert p_{4} \rvert \mathbf{3}\rangle 
    \nonumber\\
    &+\Big(
        -\langle\mathbf{2}\mathbf{3}\rangle \langle\mathbf{2}\mathbf{4}\rangle \lbrack\mathbf{1}\mathbf{4}\rbrack 
        +\langle\mathbf{2}\mathbf{4}\rangle \lbrack\mathbf{1}\mathbf{4}\rbrack \lbrack\mathbf{2}\mathbf{3}\rbrack 
        -\langle\mathbf{1}\mathbf{4}\rangle \langle\mathbf{2}\mathbf{3}\rangle \lbrack\mathbf{2}\mathbf{4}\rbrack 
        +\langle\mathbf{2}\mathbf{3}\rangle \lbrack\mathbf{1}\mathbf{4}\rbrack \lbrack\mathbf{2}\mathbf{4}\rbrack 
    \Big)\lbrack\mathbf{3}\lvert p_{4} \rvert \mathbf{1}\rangle
\Bigg) .
\label{eq:WWWW:UA full}
\end{align}

The contributions coming from the $Z$ boson in the T and U channels are 
\begin{align}
\mathcal{M}_{TZ} &= \frac{e^2}{2M_W^2M_Zs_W^2\left(t-M_Z^2\right)}
\Bigg(
    \Big(
        4c_W^{2} \langle\mathbf{2}\mathbf{4}\rangle \langle\mathbf{3}\mathbf{4}\rangle \lbrack\mathbf{1}\mathbf{2}\rbrack \lbrack\mathbf{1}\mathbf{3}\rbrack 
        -3c_W^{2} \langle\mathbf{1}\mathbf{3}\rangle \langle\mathbf{2}\mathbf{3}\rangle \langle\mathbf{2}\mathbf{4}\rangle \lbrack\mathbf{1}\mathbf{4}\rbrack 
        +4c_W^{2} \langle\mathbf{2}\mathbf{3}\rangle \langle\mathbf{2}\mathbf{4}\rangle \lbrack\mathbf{1}\mathbf{3}\rbrack \lbrack\mathbf{1}\mathbf{4}\rbrack 
        \nonumber\\
        &+3c_W^{2} \langle\mathbf{1}\mathbf{4}\rangle \langle\mathbf{2}\mathbf{4}\rangle \lbrack\mathbf{1}\mathbf{3}\rbrack \lbrack\mathbf{2}\mathbf{3}\rbrack 
        +\langle\mathbf{1}\mathbf{3}\rangle \langle\mathbf{2}\mathbf{4}\rangle \lbrack\mathbf{1}\mathbf{4}\rbrack \lbrack\mathbf{2}\mathbf{3}\rbrack 
        +3c_W^{2} \langle\mathbf{1}\mathbf{3}\rangle \langle\mathbf{2}\mathbf{4}\rangle \lbrack\mathbf{1}\mathbf{4}\rbrack \lbrack\mathbf{2}\mathbf{3}\rbrack 
        -c_W^{2} \langle\mathbf{2}\mathbf{4}\rangle \lbrack\mathbf{1}\mathbf{3}\rbrack \lbrack\mathbf{1}\mathbf{4}\rbrack \lbrack\mathbf{2}\mathbf{3}\rbrack 
        \nonumber\\
        &-3c_W^{2} \langle\mathbf{1}\mathbf{3}\rangle \langle\mathbf{1}\mathbf{4}\rangle \langle\mathbf{2}\mathbf{3}\rangle \lbrack\mathbf{2}\mathbf{4}\rbrack 
        +\langle\mathbf{1}\mathbf{4}\rangle \langle\mathbf{2}\mathbf{3}\rangle \lbrack\mathbf{1}\mathbf{3}\rbrack \lbrack\mathbf{2}\mathbf{4}\rbrack 
        +3c_W^{2} \langle\mathbf{1}\mathbf{4}\rangle \langle\mathbf{2}\mathbf{3}\rangle \lbrack\mathbf{1}\mathbf{3}\rbrack \lbrack\mathbf{2}\mathbf{4}\rbrack 
        -2c_W^{2} \langle\mathbf{1}\mathbf{3}\rangle \langle\mathbf{2}\mathbf{4}\rangle \lbrack\mathbf{1}\mathbf{3}\rbrack \lbrack\mathbf{2}\mathbf{4}\rbrack 
        \nonumber\\
        &-2\langle\mathbf{1}\mathbf{2}\rangle \langle\mathbf{3}\mathbf{4}\rangle \lbrack\mathbf{1}\mathbf{3}\rbrack \lbrack\mathbf{2}\mathbf{4}\rbrack 
        +2c_W^{2} \langle\mathbf{1}\mathbf{2}\rangle \langle\mathbf{3}\mathbf{4}\rangle \lbrack\mathbf{1}\mathbf{3}\rbrack \lbrack\mathbf{2}\mathbf{4}\rbrack 
        +c_W^{2} \langle\mathbf{1}\mathbf{3}\rangle \langle\mathbf{2}\mathbf{3}\rangle \lbrack\mathbf{1}\mathbf{4}\rbrack \lbrack\mathbf{2}\mathbf{4}\rbrack 
        +4c_W^{2} \langle\mathbf{1}\mathbf{3}\rangle \langle\mathbf{1}\mathbf{4}\rangle \lbrack\mathbf{2}\mathbf{3}\rbrack \lbrack\mathbf{2}\mathbf{4}\rbrack 
        \nonumber\\
        &-c_W^{2} \langle\mathbf{1}\mathbf{4}\rangle \lbrack\mathbf{1}\mathbf{3}\rbrack \lbrack\mathbf{2}\mathbf{3}\rbrack \lbrack\mathbf{2}\mathbf{4}\rbrack 
        +4c_W^{2} \langle\mathbf{1}\mathbf{2}\rangle \langle\mathbf{1}\mathbf{3}\rangle \lbrack\mathbf{2}\mathbf{4}\rbrack \lbrack\mathbf{3}\mathbf{4}\rbrack 
    \Big)M_Z
    \nonumber\\
    &+\Big(
        3c_W\langle\mathbf{1}\mathbf{3}\rangle \langle\mathbf{2}\mathbf{4}\rangle \lbrack\mathbf{2}\mathbf{3}\rbrack 
        +3c_W\langle\mathbf{1}\mathbf{3}\rangle \langle\mathbf{2}\mathbf{3}\rangle \lbrack\mathbf{2}\mathbf{4}\rbrack 
        -c_W\langle\mathbf{2}\mathbf{3}\rangle \lbrack\mathbf{1}\mathbf{3}\rbrack \lbrack\mathbf{2}\mathbf{4}\rbrack 
        -c_W\langle\mathbf{1}\mathbf{3}\rangle \lbrack\mathbf{2}\mathbf{3}\rbrack \lbrack\mathbf{2}\mathbf{4}\rbrack 
    \Big)\lbrack\mathbf{1}\lvert p_{3} \rvert \mathbf{4}\rangle 
    \nonumber\\
    &+\Big(
        -3c_W\langle\mathbf{2}\mathbf{3}\rangle \langle\mathbf{2}\mathbf{4}\rangle \lbrack\mathbf{1}\mathbf{3}\rbrack 
        -3c_W\langle\mathbf{1}\mathbf{3}\rangle \langle\mathbf{2}\mathbf{4}\rangle \lbrack\mathbf{2}\mathbf{3}\rbrack 
        +c_W\langle\mathbf{2}\mathbf{4}\rangle \lbrack\mathbf{1}\mathbf{3}\rbrack \lbrack\mathbf{2}\mathbf{3}\rbrack 
        +c_W\langle\mathbf{2}\mathbf{3}\rangle \lbrack\mathbf{1}\mathbf{3}\rbrack \lbrack\mathbf{2}\mathbf{4}\rbrack 
    \Big)\lbrack\mathbf{4}\lvert p_{3} \rvert \mathbf{1}\rangle
\Bigg)
\label{eq:WWWW:TZ full}
\end{align}
and 
\begin{align}
\mathcal{M}_{UZ} &= \frac{e^2}{2M_W^2M_Zs_W^2\left(u-M_Z^2\right)}
\Bigg(
    \Big(
        -3c_W^{2} \langle\mathbf{1}\mathbf{4}\rangle \langle\mathbf{2}\mathbf{3}\rangle \langle\mathbf{2}\mathbf{4}\rangle \lbrack\mathbf{1}\mathbf{3}\rbrack 
        -4c_W^{2} \langle\mathbf{2}\mathbf{3}\rangle \langle\mathbf{3}\mathbf{4}\rangle \lbrack\mathbf{1}\mathbf{2}\rbrack \lbrack\mathbf{1}\mathbf{4}\rbrack 
        +4c_W^{2} \langle\mathbf{2}\mathbf{3}\rangle \langle\mathbf{2}\mathbf{4}\rangle \lbrack\mathbf{1}\mathbf{3}\rbrack \lbrack\mathbf{1}\mathbf{4}\rbrack 
        \nonumber\\
        &-3c_W^{2} \langle\mathbf{1}\mathbf{3}\rangle \langle\mathbf{1}\mathbf{4}\rangle \langle\mathbf{2}\mathbf{4}\rangle \lbrack\mathbf{2}\mathbf{3}\rbrack 
        +c_W^{2} \langle\mathbf{1}\mathbf{4}\rangle \langle\mathbf{2}\mathbf{4}\rangle \lbrack\mathbf{1}\mathbf{3}\rbrack \lbrack\mathbf{2}\mathbf{3}\rbrack 
        -2c_W^{2} \langle\mathbf{1}\mathbf{4}\rangle \langle\mathbf{2}\mathbf{3}\rangle \lbrack\mathbf{1}\mathbf{4}\rbrack \lbrack\mathbf{2}\mathbf{3}\rbrack 
        +\langle\mathbf{1}\mathbf{3}\rangle \langle\mathbf{2}\mathbf{4}\rangle \lbrack\mathbf{1}\mathbf{4}\rbrack \lbrack\mathbf{2}\mathbf{3}\rbrack 
        \nonumber\\
        &+3c_W^{2} \langle\mathbf{1}\mathbf{3}\rangle \langle\mathbf{2}\mathbf{4}\rangle \lbrack\mathbf{1}\mathbf{4}\rbrack \lbrack\mathbf{2}\mathbf{3}\rbrack 
        +2\langle\mathbf{1}\mathbf{2}\rangle \langle\mathbf{3}\mathbf{4}\rangle \lbrack\mathbf{1}\mathbf{4}\rbrack \lbrack\mathbf{2}\mathbf{3}\rbrack 
        -2c_W^{2} \langle\mathbf{1}\mathbf{2}\rangle \langle\mathbf{3}\mathbf{4}\rangle \lbrack\mathbf{1}\mathbf{4}\rbrack \lbrack\mathbf{2}\mathbf{3}\rbrack 
        +\langle\mathbf{1}\mathbf{4}\rangle \langle\mathbf{2}\mathbf{3}\rangle \lbrack\mathbf{1}\mathbf{3}\rbrack \lbrack\mathbf{2}\mathbf{4}\rbrack 
        \nonumber\\
        &+3c_W^{2} \langle\mathbf{1}\mathbf{4}\rangle \langle\mathbf{2}\mathbf{3}\rangle \lbrack\mathbf{1}\mathbf{3}\rbrack \lbrack\mathbf{2}\mathbf{4}\rbrack 
        +3c_W^{2} \langle\mathbf{1}\mathbf{3}\rangle \langle\mathbf{2}\mathbf{3}\rangle \lbrack\mathbf{1}\mathbf{4}\rbrack \lbrack\mathbf{2}\mathbf{4}\rbrack 
        -c_W^{2} \langle\mathbf{2}\mathbf{3}\rangle \lbrack\mathbf{1}\mathbf{3}\rbrack \lbrack\mathbf{1}\mathbf{4}\rbrack \lbrack\mathbf{2}\mathbf{4}\rbrack 
        +4c_W^{2} \langle\mathbf{1}\mathbf{3}\rangle \langle\mathbf{1}\mathbf{4}\rangle \lbrack\mathbf{2}\mathbf{3}\rbrack \lbrack\mathbf{2}\mathbf{4}\rbrack 
        \nonumber\\
        &-c_W^{2} \langle\mathbf{1}\mathbf{3}\rangle \lbrack\mathbf{1}\mathbf{4}\rbrack \lbrack\mathbf{2}\mathbf{3}\rbrack \lbrack\mathbf{2}\mathbf{4}\rbrack 
        -4c_W^{2} \langle\mathbf{1}\mathbf{2}\rangle \langle\mathbf{1}\mathbf{4}\rangle \lbrack\mathbf{2}\mathbf{3}\rbrack \lbrack\mathbf{3}\mathbf{4}\rbrack 
    \Big)M_Z
    \nonumber\\
    &+\Big(
        3c_W\langle\mathbf{1}\mathbf{4}\rangle \langle\mathbf{2}\mathbf{4}\rangle \lbrack\mathbf{2}\mathbf{3}\rbrack 
        -c_W\langle\mathbf{2}\mathbf{4}\rangle \lbrack\mathbf{1}\mathbf{4}\rbrack \lbrack\mathbf{2}\mathbf{3}\rbrack 
        +3c_W\langle\mathbf{1}\mathbf{4}\rangle \langle\mathbf{2}\mathbf{3}\rangle \lbrack\mathbf{2}\mathbf{4}\rbrack 
        -c_W\langle\mathbf{1}\mathbf{4}\rangle \lbrack\mathbf{2}\mathbf{3}\rbrack \lbrack\mathbf{2}\mathbf{4}\rbrack 
    \Big)\lbrack\mathbf{1}\lvert p_{4} \rvert \mathbf{3}\rangle 
    \nonumber\\
    &+\Big(
        -3c_W\langle\mathbf{2}\mathbf{3}\rangle \langle\mathbf{2}\mathbf{4}\rangle \lbrack\mathbf{1}\mathbf{4}\rbrack 
        +c_W\langle\mathbf{2}\mathbf{4}\rangle \lbrack\mathbf{1}\mathbf{4}\rbrack \lbrack\mathbf{2}\mathbf{3}\rbrack 
        -3c_W\langle\mathbf{1}\mathbf{4}\rangle \langle\mathbf{2}\mathbf{3}\rangle \lbrack\mathbf{2}\mathbf{4}\rbrack 
        +c_W\langle\mathbf{2}\mathbf{3}\rangle \lbrack\mathbf{1}\mathbf{4}\rbrack \lbrack\mathbf{2}\mathbf{4}\rbrack 
    \Big)\lbrack\mathbf{3}\lvert p_{4} \rvert \mathbf{1}\rangle
\Bigg) .
\label{eq:WWWW:UZ full}
\end{align}

\end{widetext}

\section{\label{app:ZZWW:W T HEE}T-Channel High-Energy-Growth Terms for $\mathbf{Z Z \to W \bar{W}}$}
In this appendix, we show the high-energy growth terms separately for the $W$ boson in the T channel for the process $Z Z \to W \bar{W}$.  This serves as an example for the other processes.  Since there are many canceling terms, it would be too much to show this for every diagram.  But, we calculated the same terms for every diagram and every process discussed in this paper, although we only show combined T and U diagrams in the main body of this paper.  In the case of this process, for the U-channel diagram, each term is the same magnitude but opposite sign except for the terms that do not cancel and combine to give the result in Eq.~(\ref{eq:ZZWW:0000}).

\begin{align}
\mathcal{M}_{TW}^{(-1,-1,-1,0)} &= \frac{-3c_W^{2} e^{2} \sin(\theta)}{2\sqrt{2}M_Ws_W^{2} } \mathcal{E}\\
\mathcal{M}_{TW}^{(-1,-1,0,-1)} &= \frac{-3c_W^{2} e^{2} \sin(\theta)}{2\sqrt{2}M_Ws_W^{2} }\mathcal{E} \\
\mathcal{M}_{TW}^{(-1,-1,0,0)} &= \frac{-3c_W^{2} e^{2}\cos(\theta)}{ M_W^{2} s_W^{2} }\mathcal{E}^{2} \\
\mathcal{M}_{TW}^{(-1,-1,0,1)} &= \frac{7c_W^{2} e^{2} \sin(\theta)}{2\sqrt{2}M_Ws_W^{2} } \mathcal{E}
\end{align}
\begin{align}
\mathcal{M}_{TW}^{(-1,-1,1,0)} &= \frac{7c_W^{2} e^{2} \sin(\theta)}{2\sqrt{2}M_Ws_W^{2} } \mathcal{E}\\
\mathcal{M}_{TW}^{(-1,0,-1,-1)} &= \frac{3c_W^{3} e^{2} \sin(\theta)}{2\sqrt{2}M_Ws_W^{2} } \mathcal{E}\\
\mathcal{M}_{TW}^{(-1,0,-1,0)} &= \frac{-3c_W^{3} e^{2}\sin^{2}\left(\frac{\theta}{2}\right)}{ M_W^{2} s_W^{2}} \mathcal{E}^{2}  \\
\mathcal{M}_{TW}^{(-1,0,0,-1)} &= \frac{3c_W^{3} e^{2}\cos^{2}\left(\frac{\theta}{2}\right)}{ M_W^{2} s_W^{2}}  \mathcal{E}^{2} 
\end{align}
\begin{align}
\mathcal{M}_{TW}^{(-1,0,0,0)} &= \frac{\left(1-2c_W^{2} -c_W^{4} \right)e^{2} \sin(\theta)}{2\sqrt{2}c_WM_Ws_W^{2} } \mathcal{E}
\nonumber\\
&+\frac{-3c_W^{3} e^{2}\sin(\theta)}{\sqrt{2} M_W^{3} s_W^{2}} \mathcal{E}^{3} \\
\mathcal{M}_{TW}^{(-1,0,0,1)} &= \frac{4c_W^{3} e^{2}\sin^{2}\left(\frac{\theta}{2}\right)}{ M_W^{2} s_W^{2}} \mathcal{E}^{2}  \\
\mathcal{M}_{TW}^{(-1,0,1,0)} &= \frac{-4c_W^{3} e^{2}\cos^{2}\left(\frac{\theta}{2}\right)}{ M_W^{2} s_W^{2}}  \mathcal{E}^{2} 
\end{align}
\begin{align}
\mathcal{M}_{TW}^{(-1,0,1,1)} &= \frac{5c_W^{3} e^{2} \sin(\theta)}{2\sqrt{2}M_Ws_W^{2} } \mathcal{E}\\
\mathcal{M}_{TW}^{(0,-1,-1,-1)} &= \frac{3c_W^{3} e^{2} \sin(\theta)}{2\sqrt{2}M_Ws_W^{2} } \mathcal{E}\\
\mathcal{M}_{TW}^{(0,-1,-1,0)} &= \frac{3c_W^{3} e^{2}\cos^{2}\left(\frac{\theta}{2}\right)}{ M_W^{2} s_W^{2}}  \mathcal{E}^{2} \\
\mathcal{M}_{TW}^{(0,-1,0,-1)} &= \frac{-3c_W^{3} e^{2}\sin^{2}\left(\frac{\theta}{2}\right)}{ M_W^{2} s_W^{2}} \mathcal{E}^{2}  
\end{align}
\begin{align}
\mathcal{M}_{TW}^{(0,-1,0,0)} &= \frac{\left(1-2c_W^{2} -c_W^{4} \right)e^{2} \sin(\theta)}{2\sqrt{2}c_WM_Ws_W^{2} } \mathcal{E}
\nonumber\\
&+\frac{-3c_W^{3} e^{2}\sin(\theta)}{\sqrt{2}  M_W^{3} s_W^{2}} \mathcal{E}^{3} \\
\mathcal{M}_{TW}^{(0,-1,0,1)} &= \frac{-4c_W^{3} e^{2}\cos^{2}\left(\frac{\theta}{2}\right)}{ M_W^{2} s_W^{2}}  \mathcal{E}^{2} \\
\mathcal{M}_{TW}^{(0,-1,1,0)} &= \frac{4c_W^{3} e^{2}\sin^{2}\left(\frac{\theta}{2}\right) }{ M_W^{2} s_W^{2}} \mathcal{E}^{2} 
\end{align}
\begin{align}
\mathcal{M}_{TW}^{(0,-1,1,1)} &= \frac{5c_W^{3} e^{2} \sin(\theta)}{2\sqrt{2}M_Ws_W^{2} } \mathcal{E}\\
\mathcal{M}_{TW}^{(0,0,-1,-1)} &= \frac{-3c_W^{4} e^{2}\cos(\theta)}{M_W^{2} s_W^{2}} \mathcal{E}^{2} \\
\mathcal{M}_{TW}^{(0,0,-1,0)} &= \frac{\left(-1+5c_W^{2} +2c_W^{4} \right)e^{2} \sin(\theta)}{2\sqrt{2}M_Ws_W^{2} } \mathcal{E}
\nonumber\\
&+\frac{3c_W^{4} e^{2}\sin(\theta)}{\sqrt{2}  M_W^{3} s_W^{2}} \mathcal{E}^{3} \\
\mathcal{M}_{TW}^{(0,0,0,-1)} &= \frac{\left(-1+5c_W^{2} +2c_W^{4} \right)e^{2} \sin(\theta)}{2\sqrt{2}M_Ws_W^{2} } \mathcal{E}
\nonumber\\
&+\frac{3c_W^{4} e^{2}\sin(\theta)}{\sqrt{2}  M_W^{3} s_W^{2}} \mathcal{E}^{3} 
\end{align}
\begin{align}
\mathcal{M}_{TW}^{(0,0,0,0)} &= \frac{\left(1+\left(-1+8c_W^{2} +4c_W^{4} \right)\cos(\theta)\right)e^{2}}{2 M_W^{2} s_W^{2}} \mathcal{E}^{2} \\
\mathcal{M}_{TW}^{(0,0,0,1)} &= \frac{\left(1+-7c_W^{2} +2c_W^{4} \right)e^{2}\sin(\theta)}{2\sqrt{2}M_Ws_W^{2} }  \mathcal{E}
\nonumber\\
&+\frac{-c_W^{4} e^{2}\sin(\theta)}{\sqrt{2} M_W^{3} s_W^{2}} \mathcal{E}^{3} \\
\mathcal{M}_{TW}^{(0,0,1,0)} &= \frac{\left(1-7c_W^{2} +2c_W^{4} \right)e^{2}\sin(\theta)}{2\sqrt{2}M_Ws_W^{2} }  \mathcal{E}
\nonumber\\
&+\frac{-c_W^{4} e^{2}\sin(\theta)}{\sqrt{2}  M_W^{3} s_W^{2}} \mathcal{E}^{3} \\
\mathcal{M}_{TW}^{(0,0,1,1)} &= \frac{-c_W^{4} e^{2}\cos(\theta)}{ M_W^{2} s_W^{2}} \mathcal{E}^{2} 
\end{align}
\begin{align}
\mathcal{M}_{TW}^{(0,1,-1,-1)} &= \frac{-7c_W^{3} e^{2} \sin(\theta)}{2\sqrt{2}M_Ws_W^{2} } \mathcal{E}\\
\mathcal{M}_{TW}^{(0,1,-1,0)} &= \frac{4c_W^{3} e^{2} \sin^{2}\left(\frac{\theta}{2}\right)}{M_W^{2} s_W^{2}} \mathcal{E}^{2}  \\
\mathcal{M}_{TW}^{(0,1,0,-1)} &= \frac{-4c_W^{3} e^{2}\cos^{2}\left(\frac{\theta}{2}\right)}{ M_W^{2} s_W^{2}}  \mathcal{E}^{2} \\
\mathcal{M}_{TW}^{(0,1,0,0)} &= \frac{\left(-1+2c_W^{2} +3c_W^{4} \right)e^{2} \sin(\theta)}{2\sqrt{2}c_WM_Ws_W^{2} } \mathcal{E}
\nonumber\\
&+\frac{c_W^{3} e^{2}\sin(\theta)}{\sqrt{2} M_W^{3} s_W^{2}} \mathcal{E}^{3} 
\end{align}
\begin{align}
\mathcal{M}_{TW}^{(0,1,0,1)} &= \frac{-c_W^{3} e^{2}\sin^{2}\left(\frac{\theta}{2}\right)}{ M_W^{2} s_W^{2}} \mathcal{E}^{2}  \\
\mathcal{M}_{TW}^{(0,1,1,0)} &= \frac{c_W^{3} e^{2}\cos^{2}\left(\frac{\theta}{2}\right)}{ M_W^{2} s_W^{2}}  \mathcal{E}^{2} \\
\mathcal{M}_{TW}^{(0,1,1,1)} &= \frac{-c_W^{3} e^{2} \sin(\theta)}{2\sqrt{2}M_Ws_W^{2} } \mathcal{E}\\
\mathcal{M}_{TW}^{(1,0,-1,-1)} &= \frac{-7c_W^{3} e^{2} \sin(\theta)}{2\sqrt{2}M_Ws_W^{2} } \mathcal{E}
\end{align}
\begin{align}
\mathcal{M}_{TW}^{(1,0,-1,0)} &= \frac{-4c_W^{3} e^{2}\cos^{2}\left(\frac{\theta}{2}\right)}{ M_W^{2} s_W^{2}}  \mathcal{E}^{2} \\
\mathcal{M}_{TW}^{(1,0,0,-1)} &= \frac{4c_W^{3} e^{2}\sin^{2}\left(\frac{\theta}{2}\right)}{ M_W^{2} s_W^{2}} \mathcal{E}^{2}  \\
\mathcal{M}_{TW}^{(1,0,0,0)} &= \frac{\left(-1+2c_W^{2} +3c_W^{4} \right)e^{2} \sin(\theta)}{2\sqrt{2}c_WM_Ws_W^{2} } \mathcal{E}
\nonumber\\
&+\frac{c_W^{3} e^{2}\sin(\theta)}{\sqrt{2}M_W^{3} s_W^{2}} \mathcal{E}^{3} \\
\mathcal{M}_{TW}^{(1,0,0,1)} &= \frac{c_W^{3} e^{2}\cos^{2}\left(\frac{\theta}{2}\right)}{ M_W^{2} s_W^{2}}  \mathcal{E}^{2} 
\end{align}
\begin{align}
\mathcal{M}_{TW}^{(1,0,1,0)} &= \frac{-c_W^{3} e^{2}\sin^{2}\left(\frac{\theta}{2}\right)}{ M_W^{2} s_W^{2}} \mathcal{E}^{2}  \\
\mathcal{M}_{TW}^{(1,0,1,1)} &= \frac{-c_W^{3} e^{2} \sin(\theta)}{2\sqrt{2}M_Ws_W^{2} } \mathcal{E}\\
\mathcal{M}_{TW}^{(1,1,-1,0)} &= \frac{-5c_W^{2} e^{2} \sin(\theta)}{2\sqrt{2}M_Ws_W^{2} } \mathcal{E}\\
\mathcal{M}_{TW}^{(1,1,0,-1)} &= \frac{-5c_W^{2} e^{2} \sin(\theta)}{2\sqrt{2}M_Ws_W^{2} } \mathcal{E}
\end{align}
\begin{align}
\mathcal{M}_{TW}^{(1,1,0,0)} &= \frac{-c_W^{2} e^{2}\cos(\theta)}{ M_W^{2} s_W^{2}} \mathcal{E}^{2} \\
\mathcal{M}_{TW}^{(1,1,0,1)} &= \frac{c_W^{2} e^{2} \sin(\theta)}{2\sqrt{2}M_Ws_W^{2} } \mathcal{E}\\
\mathcal{M}_{TW}^{(1,1,1,0)} &= \frac{c_W^{2} e^{2} \sin(\theta)}{2\sqrt{2}M_Ws_W^{2} } \mathcal{E}.
\end{align}

\section{\label{app:HEG Feynman}High-Energy Growth of the Feynman 4-Point Vertex}

The polarization vectors are given by
\begin{align}
\epsilon^{(+1)\mu} &=
    \frac{1}{\sqrt{2}}\left(
\begin{array}{c}
 0 \\
 -\cos\theta \cos\phi + i \sin\phi \\
 -\cos\theta \sin\phi - i \cos\phi \\
 \sin\theta \\
\end{array}
\right)
\\
\epsilon^{(0)\mu} &= 
    \frac{1}{M}\left(\begin{array}{c}
        \lvert\mathbf{p}\rvert \\
        E\sin\theta\cos\phi\\
        E\sin\theta\sin\phi\\
        E\cos\theta
    \end{array}\right)
\\
\epsilon^{(-1)\mu} &=
    \frac{1}{\sqrt{2}}\left(
\begin{array}{c}
 0 \\
 \cos\theta \cos\phi + i \sin\phi \\
 \cos\theta \sin\phi - i \cos\phi \\
 -\sin\theta \\
\end{array}
\right).
\end{align}

\subsection{\label{app:HEG Feynman:ZZWW}$\mathbf{Z Z \to W \bar{W}}$}

The contribution from the Feynman 4-point vertex to the process $Z Z \to W \bar{W}$ is
\begin{equation}
    \mathcal{M}_{4Feyn} =
    -\frac{e^2c_W^2}{s_W^2}\Big(
    2\epsilon_1\cdot \epsilon_2 \epsilon_3^*\cdot \epsilon_4^*
    -\epsilon_1\cdot \epsilon_3^* \epsilon_2\cdot \epsilon_4^*
    -\epsilon_1\cdot \epsilon_4^* \epsilon_2\cdot \epsilon_3^*
    \Big) .
\end{equation}

If we expand this in high energy and only keep the high-energy growth terms, we get contributions whenever any of the bosons is longitudinal.  In fact, the power of $\mathcal{E}$ is equal the number of longitudinal modes.  

\begin{align}
\mathcal{M}_{4Feyn}^{(-1,-1,-1,0)} &=
	\frac{c_{\theta} c_W^2 e^2 s_{\theta} \mathcal{E}}{\sqrt{2} M_W \
s_W^2} \\
\mathcal{M}_{4Feyn}^{(-1,-1,0,-1)} &=
	-\frac{c_{\theta} c_W^2 e^2 s_{\theta} \mathcal{E}}{\sqrt{2} M_W \
s_W^2} \\
\mathcal{M}_{4Feyn}^{(-1,-1,0,0)} &=
	\frac{c_W^2 e^2 \left(s_{\theta}^2-4\right) \mathcal{E}^2}{M_W^2 \
s_W^2} \\
\mathcal{M}_{4Feyn}^{(-1,-1,0,1)} &=
	\frac{c_{\theta} c_W^2 e^2 s_{\theta} \mathcal{E}}{\sqrt{2} M_W \
s_W^2} \\
\mathcal{M}_{4Feyn}^{(-1,-1,1,0)} &=
	-\frac{c_{\theta} c_W^2 e^2 s_{\theta} \mathcal{E}}{\sqrt{2} M_W \
s_W^2} 
\end{align}
\begin{align}
\mathcal{M}_{4Feyn}^{(-1,0,-1,-1)} &=
	\frac{c_{\theta} c_W^2 e^2 s_{\theta} \mathcal{E}}{\sqrt{2} M_Z \
s_W^2} \\
\mathcal{M}_{4Feyn}^{(-1,0,-1,0)} &=
	-\frac{c_W^2 e^2 s_{\theta}^2 \mathcal{E}^2}{M_W M_Z s_W^2} \\
\mathcal{M}_{4Feyn}^{(-1,0,-1,1)} &=
	-\frac{(c_{\theta}+1) c_W^2 e^2 s_{\theta} \mathcal{E}}{\sqrt{2} M_Z \
s_W^2} \\
\mathcal{M}_{4Feyn}^{(-1,0,0,-1)} &=
	\frac{c_W^2 e^2 s_{\theta}^2 \mathcal{E}^2}{M_W M_Z s_W^2} \\
\mathcal{M}_{4Feyn}^{(-1,0,0,0)} &=
	\frac{\sqrt{2} c_{\theta} c_W^2 e^2 s_{\theta} \
\mathcal{E}^3}{M_W^2 M_Z s_W^2} 
\end{align}
\begin{align}
\mathcal{M}_{4Feyn}^{(-1,0,0,1)} &=
	\frac{c_W^2 e^2 \mathcal{E}^2 \left(c_{\theta}-s_{\theta}^2+1\right)}{M_W M_Z s_W^2} \\
\mathcal{M}_{4Feyn}^{(-1,0,1,-1)} &=
	-\frac{(c_{\theta}-1) c_W^2 e^2 s_{\theta} \mathcal{E}}{\sqrt{2} M_Z \
s_W^2} \\
\mathcal{M}_{4Feyn}^{(-1,0,1,0)} &=
	\frac{c_W^2 e^2 \mathcal{E}^2 \left(c_{\theta}+s_{\theta}^2-1\right)}{M_W M_Z s_W^2} \\
\mathcal{M}_{4Feyn}^{(-1,0,1,1)} &=
	\frac{c_{\theta} c_W^2 e^2 s_{\theta} \mathcal{E}}{\sqrt{2} M_Z \
s_W^2} 
\end{align}
\begin{align}
\mathcal{M}_{4Feyn}^{(-1,1,-1,0)} &=
	-\frac{(c_{\theta}+1) c_W^2 e^2 s_{\theta} \mathcal{E}}{\sqrt{2} M_W \
s_W^2} \\
\mathcal{M}_{4Feyn}^{(-1,1,0,-1)} &=
	\frac{(c_{\theta}-1) c_W^2 e^2 s_{\theta} \mathcal{E}}{\sqrt{2} M_W \
s_W^2} \\
\mathcal{M}_{4Feyn}^{(-1,1,0,0)} &=
	-\frac{c_W^2 e^2 s_{\theta}^2 \mathcal{E}^2}{M_W^2 s_W^2} \\
\mathcal{M}_{4Feyn}^{(-1,1,0,1)} &=
	-\frac{(c_{\theta}+1) c_W^2 e^2 s_{\theta} \mathcal{E}}{\sqrt{2} M_W \
s_W^2} \\
\mathcal{M}_{4Feyn}^{(-1,1,1,0)} &=
	\frac{(c_{\theta}-1) c_W^2 e^2 s_{\theta} \mathcal{E}}{\sqrt{2} M_W \
s_W^2} 
\end{align}
\begin{align}
\mathcal{M}_{4Feyn}^{(0,-1,-1,-1)} &=
	\frac{c_{\theta} c_W^2 e^2 s_{\theta} \mathcal{E}}{\sqrt{2} M_Z \
s_W^2} \\
\mathcal{M}_{4Feyn}^{(0,-1,-1,0)} &=
	-\frac{c_W^2 e^2 s_{\theta}^2 \mathcal{E}^2}{M_W M_Z s_W^2} \\
\mathcal{M}_{4Feyn}^{(0,-1,-1,1)} &=
	-\frac{(c_{\theta}-1) c_W^2 e^2 s_{\theta} \mathcal{E}}{\sqrt{2} M_Z \
s_W^2} \\
\mathcal{M}_{4Feyn}^{(0,-1,0,-1)} &=
	\frac{c_W^2 e^2 s_{\theta}^2 \mathcal{E}^2}{M_W M_Z s_W^2} \\
\mathcal{M}_{4Feyn}^{(0,-1,0,0)} &=
	\frac{\sqrt{2} c_{\theta} c_W^2 e^2 s_{\theta} \
\mathcal{E}^3}{M_W^2 M_Z s_W^2} 
\end{align}
\begin{align}
\mathcal{M}_{4Feyn}^{(0,-1,0,1)} &=
	-\frac{c_W^2 e^2 \mathcal{E}^2 \left(c_{\theta}+s_{\theta}^2-1\right)}{M_W M_Z s_W^2} \\
\mathcal{M}_{4Feyn}^{(0,-1,1,-1)} &=
	-\frac{(c_{\theta}+1) c_W^2 e^2 s_{\theta} \mathcal{E}}{\sqrt{2} M_Z \
s_W^2} \\
\mathcal{M}_{4Feyn}^{(0,-1,1,0)} &=
	\frac{c_W^2 e^2 \mathcal{E}^2 \left(-c_{\theta}+s_{\theta}^2-1\right)}{M_W M_Z s_W^2} \\
\mathcal{M}_{4Feyn}^{(0,-1,1,1)} &=
	\frac{c_{\theta} c_W^2 e^2 s_{\theta} \mathcal{E}}{\sqrt{2} M_Z \
s_W^2} 
\end{align}
\begin{align}
\mathcal{M}_{4Feyn}^{(0,0,-1,-1)} &=
	-\frac{c_W^2 e^2 \left(s_{\theta}^2-4\right) \mathcal{E}^2}{M_Z^2 \
s_W^2} \\
\mathcal{M}_{4Feyn}^{(0,0,-1,0)} &=
	-\frac{\sqrt{2} c_{\theta} c_W^2 e^2 s_{\theta} \mathcal{E}^3}{M_W \
M_Z^2 s_W^2} \\
\mathcal{M}_{4Feyn}^{(0,0,-1,1)} &=
	\frac{c_W^2 e^2 s_{\theta}^2 \mathcal{E}^2}{M_Z^2 s_W^2} \\
\mathcal{M}_{4Feyn}^{(0,0,0,-1)} &=
	\frac{\sqrt{2} c_{\theta} c_W^2 e^2 s_{\theta} \mathcal{E}^3}{M_W \
M_Z^2 s_W^2} \\
\mathcal{M}_{4Feyn}^{(0,0,0,0)} &=
	\frac{2 c_W^2 e^2 \mathcal{E}^2 \left(M_W^2+M_Z^2\right)}{M_W^2 
M_Z^2 s_W^2}
    \nonumber\\
    &-\frac{2 c_W^2 e^2 \left(s_{\theta}^2+2\right) \
\mathcal{E}^4}{M_W^2 M_Z^2 s_W^2} 
\end{align}
\begin{align}
\mathcal{M}_{4Feyn}^{(0,0,0,1)} &=
	-\frac{\sqrt{2} c_{\theta} c_W^2 e^2 s_{\theta} \mathcal{E}^3}{M_W \
M_Z^2 s_W^2} \\
\mathcal{M}_{4Feyn}^{(0,0,1,-1)} &=
	\frac{c_W^2 e^2 s_{\theta}^2 \mathcal{E}^2}{M_Z^2 s_W^2} \\
\mathcal{M}_{4Feyn}^{(0,0,1,0)} &=
	\frac{\sqrt{2} c_{\theta} c_W^2 e^2 s_{\theta} \mathcal{E}^3}{M_W \
M_Z^2 s_W^2} \\
\mathcal{M}_{4Feyn}^{(0,0,1,1)} &=
	-\frac{c_W^2 e^2 \left(s_{\theta}^2-4\right) \mathcal{E}^2}{M_Z^2 \
s_W^2} 
\end{align}
\begin{align}
\mathcal{M}_{4Feyn}^{(0,1,-1,-1)} &=
	-\frac{c_{\theta} c_W^2 e^2 s_{\theta} \mathcal{E}}{\sqrt{2} M_Z \
s_W^2} \\
\mathcal{M}_{4Feyn}^{(0,1,-1,0)} &=
	\frac{c_W^2 e^2 \mathcal{E}^2 \left(-c_{\theta}+s_{\theta}^2-1\right)}{M_W M_Z s_W^2} \\
\mathcal{M}_{4Feyn}^{(0,1,-1,1)} &=
	\frac{(c_{\theta}+1) c_W^2 e^2 s_{\theta} \mathcal{E}}{\sqrt{2} M_Z \
s_W^2} \\
\mathcal{M}_{4Feyn}^{(0,1,0,-1)} &=
	-\frac{c_W^2 e^2 \mathcal{E}^2 \left(c_{\theta}+s_{\theta}^2-1\right)}{M_W M_Z s_W^2} \\
\mathcal{M}_{4Feyn}^{(0,1,0,0)} &=
	-\frac{\sqrt{2} c_{\theta} c_W^2 e^2 s_{\theta} \
\mathcal{E}^3}{M_W^2 M_Z s_W^2} 
\end{align}
\begin{align}
\mathcal{M}_{4Feyn}^{(0,1,0,1)} &=
	\frac{c_W^2 e^2 s_{\theta}^2 \mathcal{E}^2}{M_W M_Z s_W^2} \\
\mathcal{M}_{4Feyn}^{(0,1,1,-1)} &=
	\frac{(c_{\theta}-1) c_W^2 e^2 s_{\theta} \mathcal{E}}{\sqrt{2} M_Z \
s_W^2} \\
\mathcal{M}_{4Feyn}^{(0,1,1,0)} &=
	-\frac{c_W^2 e^2 s_{\theta}^2 \mathcal{E}^2}{M_W M_Z s_W^2} \\
\mathcal{M}_{4Feyn}^{(0,1,1,1)} &=
	-\frac{c_{\theta} c_W^2 e^2 s_{\theta} \mathcal{E}}{\sqrt{2} M_Z \
s_W^2} 
\end{align}
\begin{align}
\mathcal{M}_{4Feyn}^{(1,-1,-1,0)} &=
	-\frac{(c_{\theta}-1) c_W^2 e^2 s_{\theta} \mathcal{E}}{\sqrt{2} M_W \
s_W^2} \\
\mathcal{M}_{4Feyn}^{(1,-1,0,-1)} &=
	\frac{(c_{\theta}+1) c_W^2 e^2 s_{\theta} \mathcal{E}}{\sqrt{2} M_W \
s_W^2} \\
\mathcal{M}_{4Feyn}^{(1,-1,0,0)} &=
	-\frac{c_W^2 e^2 s_{\theta}^2 \mathcal{E}^2}{M_W^2 s_W^2} \\
\mathcal{M}_{4Feyn}^{(1,-1,0,1)} &=
	-\frac{(c_{\theta}-1) c_W^2 e^2 s_{\theta} \mathcal{E}}{\sqrt{2} M_W \
s_W^2} \\
\mathcal{M}_{4Feyn}^{(1,-1,1,0)} &=
	\frac{(c_{\theta}+1) c_W^2 e^2 s_{\theta} \mathcal{E}}{\sqrt{2} M_W \
s_W^2} 
\end{align}
\begin{align}
\mathcal{M}_{4Feyn}^{(1,0,-1,-1)} &=
	-\frac{c_{\theta} c_W^2 e^2 s_{\theta} \mathcal{E}}{\sqrt{2} M_Z \
s_W^2} \\
\mathcal{M}_{4Feyn}^{(1,0,-1,0)} &=
	\frac{c_W^2 e^2 \mathcal{E}^2 \left(c_{\theta}+s_{\theta}^2-1\right)}{M_W M_Z s_W^2} \\
\mathcal{M}_{4Feyn}^{(1,0,-1,1)} &=
	\frac{(c_{\theta}-1) c_W^2 e^2 s_{\theta} \mathcal{E}}{\sqrt{2} M_Z \
s_W^2} \\
\mathcal{M}_{4Feyn}^{(1,0,0,-1)} &=
	\frac{c_W^2 e^2 \mathcal{E}^2 \left(c_{\theta}-s_{\theta}^2+1\right)}{M_W M_Z s_W^2} \\
\mathcal{M}_{4Feyn}^{(1,0,0,0)} &=
	-\frac{\sqrt{2} c_{\theta} c_W^2 e^2 s_{\theta} \
\mathcal{E}^3}{M_W^2 M_Z s_W^2} 
\end{align}
\begin{align}
\mathcal{M}_{4Feyn}^{(1,0,0,1)} &=
	\frac{c_W^2 e^2 s_{\theta}^2 \mathcal{E}^2}{M_W M_Z s_W^2} \\
\mathcal{M}_{4Feyn}^{(1,0,1,-1)} &=
	\frac{(c_{\theta}+1) c_W^2 e^2 s_{\theta} \mathcal{E}}{\sqrt{2} M_Z \
s_W^2} \\
\mathcal{M}_{4Feyn}^{(1,0,1,0)} &=
	-\frac{c_W^2 e^2 s_{\theta}^2 \mathcal{E}^2}{M_W M_Z s_W^2} \\
\mathcal{M}_{4Feyn}^{(1,0,1,1)} &=
	-\frac{c_{\theta} c_W^2 e^2 s_{\theta} \mathcal{E}}{\sqrt{2} M_Z \
s_W^2} 
\end{align}
\begin{align}
\mathcal{M}_{4Feyn}^{(1,1,-1,0)} &=
	\frac{c_{\theta} c_W^2 e^2 s_{\theta} \mathcal{E}}{\sqrt{2} M_W \
s_W^2} \\
\mathcal{M}_{4Feyn}^{(1,1,0,-1)} &=
	-\frac{c_{\theta} c_W^2 e^2 s_{\theta} \mathcal{E}}{\sqrt{2} M_W \
s_W^2} \\
\mathcal{M}_{4Feyn}^{(1,1,0,0)} &=
	\frac{c_W^2 e^2 \left(s_{\theta}^2-4\right) \mathcal{E}^2}{M_W^2 \
s_W^2} \\
\mathcal{M}_{4Feyn}^{(1,1,0,1)} &=
	\frac{c_{\theta} c_W^2 e^2 s_{\theta} \mathcal{E}}{\sqrt{2} M_W \
s_W^2} \\
\mathcal{M}_{4Feyn}^{(1,1,1,0)} &=
	-\frac{c_{\theta} c_W^2 e^2 s_{\theta} \mathcal{E}}{\sqrt{2} M_W \
s_W^2} 
.
\end{align}

\subsection{\label{app:HEG Feynman:WWWW}$\mathbf{W W \to W W}$}

The contribution from the Feynman 4-point vertex to the process $W W \to W W$ is
\begin{equation}
    \mathcal{M}_{4Feyn} =
    \frac{e^2}{s_W^2}\Big(
    2\epsilon_1\cdot \epsilon_4^* \epsilon_2\cdot\epsilon_3^*
    -\epsilon_1\cdot \epsilon_3^* \epsilon_2\cdot \epsilon_4^*
    -\epsilon_1\cdot \epsilon_2 \epsilon_3^*\cdot\epsilon_4^*
    \Big) .
\end{equation}

If we expand this in high energy and only keep the high-energy growth terms, we get contributions whenever any of the bosons is longitudinal.  Once again, the power of $\mathcal{E}$ is equal the number of longitudinal modes.  

\begin{align}
\mathcal{M}_{4Feyn}^{(-1,-1,-1,0)} &=
	\frac{(c_{\theta}-3) e^2 s_{\theta} \mathcal{E}}{2 \sqrt{2} M_W s_W^2} \\
\mathcal{M}_{4Feyn}^{(-1,-1,0,-1)} &=
	-\frac{(c_{\theta}-3) e^2 s_{\theta} \mathcal{E}}{2 \sqrt{2} M_W s_W^2} \\
\mathcal{M}_{4Feyn}^{(-1,-1,0,0)} &=
	\frac{e^2 s_{\theta}^2 \mathcal{E}^2}{2 M_W^2 s_W^2} \\
\mathcal{M}_{4Feyn}^{(-1,-1,0,1)} &=
	\frac{(c_{\theta}-1) e^2 s_{\theta} \mathcal{E}}{2 \sqrt{2} M_W s_W^2} \\
\mathcal{M}_{4Feyn}^{(-1,-1,1,0)} &=
	-\frac{(c_{\theta}-1) e^2 s_{\theta} \mathcal{E}}{2 \sqrt{2} M_W s_W^2} 
\end{align}
\begin{align}
\mathcal{M}_{4Feyn}^{(-1,0,-1,-1)} &=
	\frac{(c_{\theta}-3) e^2 s_{\theta} \mathcal{E}}{2 \sqrt{2} M_W s_W^2} \\
\mathcal{M}_{4Feyn}^{(-1,0,-1,0)} &=
	-\frac{e^2 s_{\theta}^2 \mathcal{E}^2}{2 M_W^2 s_W^2} \\
\mathcal{M}_{4Feyn}^{(-1,0,-1,1)} &=
	-\frac{(c_{\theta}+1) e^2 s_{\theta} \mathcal{E}}{2 \sqrt{2} M_W s_W^2} \\
\mathcal{M}_{4Feyn}^{(-1,0,0,-1)} &=
	\frac{e^2 \left(s_{\theta}^2-4\right) \mathcal{E}^2}{2 M_W^2 s_W^2} \\
\mathcal{M}_{4Feyn}^{(-1,0,0,0)} &=
	\frac{(c_{\theta}+3) e^2 s_{\theta} \mathcal{E}^3}{\sqrt{2} M_W^3 s_W^2}-\frac{3 e^2 s_{\theta} \mathcal{E}}{\sqrt{2} M_W s_W^2} 
\end{align}
\begin{align}
\mathcal{M}_{4Feyn}^{(-1,0,0,1)} &=
	-\frac{e^2 s_{\theta}^2 \mathcal{E}^2}{2 M_W^2 s_W^2} \\
\mathcal{M}_{4Feyn}^{(-1,0,1,-1)} &=
	-\frac{(c_{\theta}+3) e^2 s_{\theta} \mathcal{E}}{2 \sqrt{2} M_W s_W^2} \\
\mathcal{M}_{4Feyn}^{(-1,0,1,0)} &=
	\frac{e^2 \mathcal{E}^2 \left(-2 c_{\theta}+s_{\theta}^2+2\right)}{2 M_W^2 s_W^2} \\
\mathcal{M}_{4Feyn}^{(-1,0,1,1)} &=
	\frac{(c_{\theta}-1) e^2 s_{\theta} \mathcal{E}}{2 \sqrt{2} M_W s_W^2} 
\end{align}
\begin{align}
\mathcal{M}_{4Feyn}^{(-1,1,-1,0)} &=
	-\frac{(c_{\theta}+1) e^2 s_{\theta} \mathcal{E}}{2 \sqrt{2} M_W s_W^2} \\
\mathcal{M}_{4Feyn}^{(-1,1,0,-1)} &=
	\frac{(c_{\theta}+3) e^2 s_{\theta} \mathcal{E}}{2 \sqrt{2} M_W s_W^2} \\
\mathcal{M}_{4Feyn}^{(-1,1,0,0)} &=
	\frac{e^2 \mathcal{E}^2 \left(4 c_{\theta}-s_{\theta}^2+4\right)}{2 M_W^2 s_W^2} \\
\mathcal{M}_{4Feyn}^{(-1,1,0,1)} &=
	-\frac{(c_{\theta}+1) e^2 s_{\theta} \mathcal{E}}{2 \sqrt{2} M_W s_W^2} \\
\mathcal{M}_{4Feyn}^{(-1,1,1,0)} &=
	\frac{(c_{\theta}+3) e^2 s_{\theta} \mathcal{E}}{2 \sqrt{2} M_W s_W^2} 
\end{align}
\begin{align}
\mathcal{M}_{4Feyn}^{(0,-1,-1,-1)} &=
	\frac{(c_{\theta}-3) e^2 s_{\theta} \mathcal{E}}{2 \sqrt{2} M_W s_W^2} \\
\mathcal{M}_{4Feyn}^{(0,-1,-1,0)} &=
	-\frac{e^2 \left(s_{\theta}^2-4\right) \mathcal{E}^2}{2 M_W^2 s_W^2} \\
\mathcal{M}_{4Feyn}^{(0,-1,-1,1)} &=
	-\frac{(c_{\theta}+3) e^2 s_{\theta} \mathcal{E}}{2 \sqrt{2} M_W s_W^2} \\
\mathcal{M}_{4Feyn}^{(0,-1,0,-1)} &=
	\frac{e^2 s_{\theta}^2 \mathcal{E}^2}{2 M_W^2 s_W^2} \\
\mathcal{M}_{4Feyn}^{(0,-1,0,0)} &=
	\frac{(c_{\theta}+3) e^2 s_{\theta} \mathcal{E}^3}{\sqrt{2} M_W^3 s_W^2}-\frac{3 e^2 s_{\theta} \mathcal{E}}{\sqrt{2} M_W s_W^2} 
\end{align}
\begin{align}
\mathcal{M}_{4Feyn}^{(0,-1,0,1)} &=
	\frac{e^2 \mathcal{E}^2 \left(2 c_{\theta}-s_{\theta}^2-2\right)}{2 M_W^2 s_W^2} \\
\mathcal{M}_{4Feyn}^{(0,-1,1,-1)} &=
	-\frac{(c_{\theta}+1) e^2 s_{\theta} \mathcal{E}}{2 \sqrt{2} M_W s_W^2} \\
\mathcal{M}_{4Feyn}^{(0,-1,1,0)} &=
	\frac{e^2 s_{\theta}^2 \mathcal{E}^2}{2 M_W^2 s_W^2} \\
\mathcal{M}_{4Feyn}^{(0,-1,1,1)} &=
	\frac{(c_{\theta}-1) e^2 s_{\theta} \mathcal{E}}{2 \sqrt{2} M_W s_W^2} 
\end{align}
\begin{align}
\mathcal{M}_{4Feyn}^{(0,0,-1,-1)} &=
	-\frac{e^2 s_{\theta}^2 \mathcal{E}^2}{2 M_W^2 s_W^2} \\
\mathcal{M}_{4Feyn}^{(0,0,-1,0)} &=
	\frac{3 e^2 s_{\theta} \mathcal{E}}{\sqrt{2} M_W s_W^2}-\frac{(c_{\theta}+3) e^2 s_{\theta} \mathcal{E}^3}{\sqrt{2} M_W^3 s_W^2} \\
\mathcal{M}_{4Feyn}^{(0,0,-1,1)} &=
	\frac{e^2 \mathcal{E}^2 \left(-4 c_{\theta}+s_{\theta}^2-4\right)}{2 M_W^2 s_W^2} \\
\mathcal{M}_{4Feyn}^{(0,0,0,-1)} &=
	\frac{(c_{\theta}+3) e^2 s_{\theta} \mathcal{E}^3}{\sqrt{2} M_W^3 s_W^2}-\frac{3 e^2 s_{\theta} \mathcal{E}}{\sqrt{2} M_W s_W^2} \\
\mathcal{M}_{4Feyn}^{(0,0,0,0)} &=
	\frac{e^2 \mathcal{E}^4 \left(6 c_{\theta}-s_{\theta}^2-2\right)}{M_W^4 s_W^2}
 \nonumber\\
 &+\frac{2 (1-3 c_{\theta}) e^2 \mathcal{E}^2}{M_W^2 s_W^2} 
\end{align}
\begin{align}
\mathcal{M}_{4Feyn}^{(0,0,0,1)} &=
	\frac{3 e^2 s_{\theta} \mathcal{E}}{\sqrt{2} M_W s_W^2}-\frac{(c_{\theta}+3) e^2 s_{\theta} \mathcal{E}^3}{\sqrt{2} M_W^3 s_W^2} \\
\mathcal{M}_{4Feyn}^{(0,0,1,-1)} &=
	\frac{e^2 \mathcal{E}^2 \left(-4 c_{\theta}+s_{\theta}^2-4\right)}{2 M_W^2 s_W^2} \\
\mathcal{M}_{4Feyn}^{(0,0,1,0)} &=
	\frac{(c_{\theta}+3) e^2 s_{\theta} \mathcal{E}^3}{\sqrt{2} M_W^3 s_W^2}-\frac{3 e^2 s_{\theta} \mathcal{E}}{\sqrt{2} M_W s_W^2} \\
\mathcal{M}_{4Feyn}^{(0,0,1,1)} &=
	-\frac{e^2 s_{\theta}^2 \mathcal{E}^2}{2 M_W^2 s_W^2} 
\end{align}
\begin{align}
\mathcal{M}_{4Feyn}^{(0,1,-1,-1)} &=
	-\frac{(c_{\theta}-1) e^2 s_{\theta} \mathcal{E}}{2 \sqrt{2} M_W s_W^2} \\
\mathcal{M}_{4Feyn}^{(0,1,-1,0)} &=
	\frac{e^2 s_{\theta}^2 \mathcal{E}^2}{2 M_W^2 s_W^2} \\
\mathcal{M}_{4Feyn}^{(0,1,-1,1)} &=
	\frac{(c_{\theta}+1) e^2 s_{\theta} \mathcal{E}}{2 \sqrt{2} M_W s_W^2} \\
\mathcal{M}_{4Feyn}^{(0,1,0,-1)} &=
	\frac{e^2 \mathcal{E}^2 \left(2 c_{\theta}-s_{\theta}^2-2\right)}{2 M_W^2 s_W^2} \\
\mathcal{M}_{4Feyn}^{(0,1,0,0)} &=
	\frac{3 e^2 s_{\theta} \mathcal{E}}{\sqrt{2} M_W s_W^2}-\frac{(c_{\theta}+3) e^2 s_{\theta} \mathcal{E}^3}{\sqrt{2} M_W^3 s_W^2} 
\end{align}
\begin{align}
\mathcal{M}_{4Feyn}^{(0,1,0,1)} &=
	\frac{e^2 s_{\theta}^2 \mathcal{E}^2}{2 M_W^2 s_W^2} \\
\mathcal{M}_{4Feyn}^{(0,1,1,-1)} &=
	\frac{(c_{\theta}+3) e^2 s_{\theta} \mathcal{E}}{2 \sqrt{2} M_W s_W^2} \\
\mathcal{M}_{4Feyn}^{(0,1,1,0)} &=
	-\frac{e^2 \left(s_{\theta}^2-4\right) \mathcal{E}^2}{2 M_W^2 s_W^2} \\
\mathcal{M}_{4Feyn}^{(0,1,1,1)} &=
	-\frac{(c_{\theta}-3) e^2 s_{\theta} \mathcal{E}}{2 \sqrt{2} M_W s_W^2} 
\end{align}
\begin{align}
\mathcal{M}_{4Feyn}^{(1,-1,-1,0)} &=
	-\frac{(c_{\theta}+3) e^2 s_{\theta} \mathcal{E}}{2 \sqrt{2} M_W s_W^2} \\
\mathcal{M}_{4Feyn}^{(1,-1,0,-1)} &=
	\frac{(c_{\theta}+1) e^2 s_{\theta} \mathcal{E}}{2 \sqrt{2} M_W s_W^2} \\
\mathcal{M}_{4Feyn}^{(1,-1,0,0)} &=
	\frac{e^2 \mathcal{E}^2 \left(4 c_{\theta}-s_{\theta}^2+4\right)}{2 M_W^2 s_W^2} \\
\mathcal{M}_{4Feyn}^{(1,-1,0,1)} &=
	-\frac{(c_{\theta}+3) e^2 s_{\theta} \mathcal{E}}{2 \sqrt{2} M_W s_W^2} \\
\mathcal{M}_{4Feyn}^{(1,-1,1,0)} &=
	\frac{(c_{\theta}+1) e^2 s_{\theta} \mathcal{E}}{2 \sqrt{2} M_W s_W^2} 
\end{align}
\begin{align}
\mathcal{M}_{4Feyn}^{(1,0,-1,-1)} &=
	-\frac{(c_{\theta}-1) e^2 s_{\theta} \mathcal{E}}{2 \sqrt{2} M_W s_W^2} \\
\mathcal{M}_{4Feyn}^{(1,0,-1,0)} &=
	\frac{e^2 \mathcal{E}^2 \left(-2 c_{\theta}+s_{\theta}^2+2\right)}{2 M_W^2 s_W^2} \\
\mathcal{M}_{4Feyn}^{(1,0,-1,1)} &=
	\frac{(c_{\theta}+3) e^2 s_{\theta} \mathcal{E}}{2 \sqrt{2} M_W s_W^2} \\
\mathcal{M}_{4Feyn}^{(1,0,0,-1)} &=
	-\frac{e^2 s_{\theta}^2 \mathcal{E}^2}{2 M_W^2 s_W^2} \\
\mathcal{M}_{4Feyn}^{(1,0,0,0)} &=
	\frac{3 e^2 s_{\theta} \mathcal{E}}{\sqrt{2} M_W s_W^2}-\frac{(c_{\theta}+3) e^2 s_{\theta} \mathcal{E}^3}{\sqrt{2} M_W^3 s_W^2} 
\end{align}
\begin{align}
\mathcal{M}_{4Feyn}^{(1,0,0,1)} &=
	\frac{e^2 \left(s_{\theta}^2-4\right) \mathcal{E}^2}{2 M_W^2 s_W^2} \\
\mathcal{M}_{4Feyn}^{(1,0,1,-1)} &=
	\frac{(c_{\theta}+1) e^2 s_{\theta} \mathcal{E}}{2 \sqrt{2} M_W s_W^2} \\
\mathcal{M}_{4Feyn}^{(1,0,1,0)} &=
	-\frac{e^2 s_{\theta}^2 \mathcal{E}^2}{2 M_W^2 s_W^2} \\
\mathcal{M}_{4Feyn}^{(1,0,1,1)} &=
	-\frac{(c_{\theta}-3) e^2 s_{\theta} \mathcal{E}}{2 \sqrt{2} M_W s_W^2} 
\end{align}
\begin{align}
\mathcal{M}_{4Feyn}^{(1,1,-1,0)} &=
	\frac{(c_{\theta}-1) e^2 s_{\theta} \mathcal{E}}{2 \sqrt{2} M_W s_W^2} \\
\mathcal{M}_{4Feyn}^{(1,1,0,-1)} &=
	-\frac{(c_{\theta}-1) e^2 s_{\theta} \mathcal{E}}{2 \sqrt{2} M_W s_W^2} \\
\mathcal{M}_{4Feyn}^{(1,1,0,0)} &=
	\frac{e^2 s_{\theta}^2 \mathcal{E}^2}{2 M_W^2 s_W^2} \\
\mathcal{M}_{4Feyn}^{(1,1,0,1)} &=
	\frac{(c_{\theta}-3) e^2 s_{\theta} \mathcal{E}}{2 \sqrt{2} M_W s_W^2} \\
\mathcal{M}_{4Feyn}^{(1,1,1,0)} &=
	-\frac{(c_{\theta}-3) e^2 s_{\theta} \mathcal{E}}{2 \sqrt{2} M_W s_W^2}
 .
\end{align}

\subsection{\label{app:HEG Feynman:hhZZ}$\mathbf{h h \to Z Z}$ and $\mathbf{h h \to W \bar{W}}$}

The contribution from the Feynman 4-point vertex to the process $h h \to Z Z$ is
\begin{equation}
    \mathcal{M}_{4Feyn} =
    -\frac{e^2}{2c_W^2s_W^2}\epsilon_3^*\cdot \epsilon_4^* .
\end{equation}

The contribution from the Feynman 4-point vertex to the process $h h \to W \bar{W}$ is
\begin{equation}
    \mathcal{M}_{4Feyn} =
    -\frac{e^2}{2s_W^2}\epsilon_3^*\cdot \epsilon_4^* .
\end{equation}

If we expand these in high energy and only keep the high-energy growth terms, we get the following nonzero contributions.  The result is the same for both processes, namely,
\begin{align}
\mathcal{M}_{4Feyn}^{(0,0)} &=
	-\frac{e^2 \mathcal{E}^2}{M_W^2 s_W^2}
 .
\end{align}

\subsection{\label{app:HEG Feynman:AZWW}$\mathbf{\gamma^+ Z \to W \bar{W}}$}

The contribution from the Feynman 4-point vertex to the process $\gamma^+ Z \to W \bar{W}$ is
\begin{equation}
    \mathcal{M}_{4Feyn} =
    -\frac{e^2c_W}{s_W}\Big(
    2\epsilon_1\cdot \epsilon_4^* \epsilon_2\cdot\epsilon_3^*
    -\epsilon_1\cdot \epsilon_3^* \epsilon_2\cdot \epsilon_4^*
    -\epsilon_1\cdot \epsilon_2 \epsilon_3^*\cdot\epsilon_4^*
    \Big) .
\end{equation}

If we expand this in high energy and only keep the high-energy growth terms, we get contributions whenever any of the bosons is longitudinal. 

\begin{align}
\mathcal{M}_{4Feyn}^{(1,-1,-1,0)} &=
	\frac{c_W e^2 s_{\theta} \mathcal{E} (2 (c_{\theta}+1) M_W+(c_{\theta}-1) M_Z)}{2 \sqrt{2} M_W M_Z s_W} \\
\mathcal{M}_{4Feyn}^{(1,-1,0,-1)} &=
	\frac{(c_{\theta}+1) c_W e^2 s_{\theta} \mathcal{E}}{2 \sqrt{2} M_W s_W} \\
\mathcal{M}_{4Feyn}^{(1,-1,0,0)} &=
	\frac{c_W e^2 \mathcal{E}^2 \left(M_W \left(2 s_{\theta}^2-4 c_{\theta}-4\right)+M_Z s_{\theta}^2\right)}{2 M_W^2 M_Z s_W} \\
\mathcal{M}_{4Feyn}^{(1,-1,0,1)} &=
	-\frac{(c_{\theta}+3) c_W e^2 s_{\theta} \mathcal{E}}{2 \sqrt{2} M_W s_W} \\
\mathcal{M}_{4Feyn}^{(1,-1,1,0)} &=
	-\frac{(c_{\theta}+1) c_W e^2 s_{\theta} \mathcal{E} (2 M_W+M_Z)}{2 \sqrt{2} M_W M_Z s_W} 
\end{align}
\begin{align}
\mathcal{M}_{4Feyn}^{(1,0,-1,-1)} &=
	-\frac{(c_{\theta}-1) c_W e^2 s_{\theta} \mathcal{E} (M_W+2 M_Z)}{2 \sqrt{2} M_W M_Z s_W} \\
\mathcal{M}_{4Feyn}^{(1,0,-1,0)} &=
	\frac{c_W e^2 \mathcal{E}^2 \left(2 c_{\theta}-s_{\theta}^2-2\right)}{2 M_W M_Z s_W} \\
\mathcal{M}_{4Feyn}^{(1,0,-1,1)} &=
	\frac{c_W e^2 s_{\theta} \mathcal{E} ((c_{\theta}-1) M_W+2 (c_{\theta}+1) M_Z)}{2 \sqrt{2} M_W M_Z s_W} \\
\mathcal{M}_{4Feyn}^{(1,0,0,-1)} &=
	-\frac{c_W e^2 s_{\theta}^2 \mathcal{E}^2 (M_W+2 M_Z)}{2 M_W^2 M_Z s_W} \\
\mathcal{M}_{4Feyn}^{(1,0,0,0)} &=
	\frac{(c_{\theta}+3) c_W e^2 s_{\theta} \mathcal{E}^3}{\sqrt{2} M_W^2 M_Z s_W}
    \nonumber\\
    &-\frac{c_W e^2 s_{\theta} \mathcal{E} \left(6 M_W^2-(c_{\theta}-3) M_Z^2\right)}{4 \sqrt{2} M_W^2 M_Z s_W} 
\end{align}
\begin{align}
\mathcal{M}_{4Feyn}^{(1,0,0,1)} &=
	\frac{c_W e^2 \mathcal{E}^2 \left(M_W s_{\theta}^2+2 M_Z \left(s_{\theta}^2-2\right)\right)}{2 M_W^2 M_Z s_W} \\
\mathcal{M}_{4Feyn}^{(1,0,1,-1)} &=
	\frac{(c_{\theta}+1) c_W e^2 s_{\theta} \mathcal{E} (M_W+2 M_Z)}{2 \sqrt{2} M_W M_Z s_W} \\
\mathcal{M}_{4Feyn}^{(1,0,1,0)} &=
	\frac{c_W e^2 s_{\theta}^2 \mathcal{E}^2}{2 M_W M_Z s_W} \\
\mathcal{M}_{4Feyn}^{(1,0,1,1)} &=
	-\frac{c_W e^2 s_{\theta} \mathcal{E} ((c_{\theta}+1) M_W+2 (c_{\theta}-1) M_Z)}{2 \sqrt{2} M_W M_Z s_W}
\end{align}
\begin{align}
\mathcal{M}_{4Feyn}^{(1,1,-1,0)} &=
	-\frac{(c_{\theta}-1) c_W e^2 s_{\theta} \mathcal{E} (2 M_W+M_Z)}{2 \sqrt{2} M_W M_Z s_W} \\
\mathcal{M}_{4Feyn}^{(1,1,0,-1)} &=
	-\frac{(c_{\theta}-1) c_W e^2 s_{\theta} \mathcal{E}}{2 \sqrt{2} M_W s_W} \\
\mathcal{M}_{4Feyn}^{(1,1,0,0)} &=
	-\frac{c_W e^2 s_{\theta}^2 \mathcal{E}^2 (2 M_W+M_Z)}{2 M_W^2 M_Z s_W} \\
\mathcal{M}_{4Feyn}^{(1,1,0,1)} &=
	\frac{(c_{\theta}-3) c_W e^2 s_{\theta} \mathcal{E}}{2 \sqrt{2} M_W s_W} \\
\mathcal{M}_{4Feyn}^{(1,1,1,0)} &=
	\frac{c_W e^2 s_{\theta} \mathcal{E} (2 (c_{\theta}-1) M_W+(c_{\theta}+1) M_Z)}{2 \sqrt{2} M_W M_Z s_W}
 .
\end{align}

\subsection{\label{app:HEG Feynman:AAWW}$\mathbf{\gamma^+ \gamma^{\pm} \to W \bar{W}}$}

The contribution from the Feynman 4-point vertex to the process $\gamma \gamma \to W \bar{W}$ is
\begin{equation}
    \mathcal{M}_{4Feyn} =
    -e^2\Big(
    2\epsilon_1\cdot \epsilon_2 \epsilon_3^*\cdot \epsilon_4^*
    -\epsilon_1\cdot \epsilon_3^* \epsilon_2\cdot \epsilon_4^*
    -\epsilon_1\cdot \epsilon_4^* \epsilon_2\cdot \epsilon_3^*
    \Big) .
\end{equation}

We first expand this in high energy when both photons have positive helicity.

\begin{align}
\mathcal{M}_{4Feyn}^{(1,1,-1,0)} &=
	\frac{c_{\theta} e^2 s_{\theta} \mathcal{E}}{\sqrt{2} M_W} \\
\mathcal{M}_{4Feyn}^{(1,1,0,-1)} &=
	-\frac{c_{\theta} e^2 s_{\theta} \mathcal{E}}{\sqrt{2} M_W} \\
\mathcal{M}_{4Feyn}^{(1,1,0,0)} &=
	\frac{e^2 \left(s_{\theta}^2-4\right) \mathcal{E}^2}{M_W^2} \\
\mathcal{M}_{4Feyn}^{(1,1,0,1)} &=
	\frac{c_{\theta} e^2 s_{\theta} \mathcal{E}}{\sqrt{2} M_W} \\
\mathcal{M}_{4Feyn}^{(1,1,1,0)} &=
	-\frac{c_{\theta} e^2 s_{\theta} \mathcal{E}}{\sqrt{2} M_W} .
\end{align}

We next expand this when the first photon has positive helicity and the second has negative helicity.

\begin{align}
\mathcal{M}_{4Feyn}^{(1,-1,-1,0)} &=
	-\frac{(c_{\theta}-1) e^2 s_{\theta} \mathcal{E}}{\sqrt{2} M_W} \\
\mathcal{M}_{4Feyn}^{(1,-1,0,-1)} &=
	\frac{(c_{\theta}+1) e^2 s_{\theta} \mathcal{E}}{\sqrt{2} M_W} \\
\mathcal{M}_{4Feyn}^{(1,-1,0,0)} &=
	-\frac{e^2 s_{\theta}^2 \mathcal{E}^2}{M_W^2} \\
\mathcal{M}_{4Feyn}^{(1,-1,0,1)} &=
	-\frac{(c_{\theta}-1) e^2 s_{\theta} \mathcal{E}}{\sqrt{2} M_W} \\
\mathcal{M}_{4Feyn}^{(1,-1,1,0)} &=
	\frac{(c_{\theta}+1) e^2 s_{\theta} \mathcal{E}}{\sqrt{2} M_W} .
 \\
 \nonumber
\end{align}

\end{document}